\documentclass[prd,twocolumn,superscriptaddress,nofootinbib]{revtex4-2} 
\usepackage{graphicx}
\usepackage{amsmath}
\usepackage{relsize}
\usepackage{booktabs}
\usepackage{url}
\usepackage{tabularx}
\usepackage{array}
\usepackage{siunitx}
\usepackage{xcolor}
\usepackage[colorlinks=true, allcolors=blue]{hyperref}

\newcommand{\arasim}{\texttt{AraSim}}
\newcommand{\araroot}{\texttt{AraRoot}}
\newcommand{\nuleptonsim}{\texttt{NuLeptonSim}}

\newcommand{\atUC}{\affiliation{Dept.~of Physics, Dept.~of Astronomy and Astrophysics, Enrico Fermi Institute, Kavli Institute for Cosmological Physics, University of Chicago, Chicago, IL 60637}}
\newcommand{\atKU}{\affiliation{Dept.~of Physics and Astronomy, University of Kansas, Lawrence, KS 66045}}
\newcommand{\atOSU}{\affiliation{Dept.~of Physics, Center for Cosmology and AstroParticle Physics, The Ohio State University, Columbus, OH 43210}}
\newcommand{\atUW}{\affiliation{Dept.~of Physics, Wisconsin IceCube Particle Astrophysics Center, University of Wisconsin-Madison, Madison,  WI 53706}}
\newcommand{\atNTU}{\affiliation{Dept.~of Physics, Graduate Institute of Astrophysics, Leung Center for Cosmology and Particle Astrophysics, National Taiwan University, Taipei, Taiwan}}
\newcommand{\atULB}{\affiliation{Universite Libre de Bruxelles, Science Faculty CP230, B-1050 Brussels, Belgium}}
\newcommand{\atUMD}{\affiliation{Dept.~of Physics, University of Maryland, College Park, MD 20742}}
\newcommand{\atUCL}{\affiliation{Dept.~of Physics and Astronomy, University College London, London, United Kingdom}}
\newcommand{\atPSUigc}{\affiliation{Center for Multi-Messenger Astrophysics, Institute for Gravitation and the Cosmos, Pennsylvania State University, University Park, PA 16802}}
\newcommand{\atPSUphys}{\affiliation{Dept.~of Physics, Pennsylvania State University, University Park, PA 16802}}
\newcommand{\atPSUast}{\affiliation{Dept.~of Astronomy and Astrophysics, Pennsylvania State University, University Park, PA 16802}}
\newcommand{\atVUB}{\affiliation{Vrije Universiteit Brussel, Brussels, Belgium}}
\newcommand{\atChiba}{\affiliation{International Center for Hadron Astrophysics, Chiba University, Chiba 263-8522 Japan}}
\newcommand{\atUNL}{\affiliation{Dept.~of Physics and Astronomy, University of Nebraska, Lincoln, Nebraska 68588}}
\newcommand{\atWhittier}{\affiliation{Dept.~Physics and Astronomy, Whittier College, Whittier, CA 90602}}
\newcommand{\atUD}{\affiliation{Dept.~of Physics, University of Delaware, Newark, DE 19716}}
\newcommand{\atNUU}{\affiliation{Dept.~of Energy Engineering, National United University, Miaoli, Taiwan}}
\newcommand{\atNPU}{\affiliation{Dept.~of Applied Physics, National Pingtung University, Pingtung City, Pingtung County 900393, Taiwan}}
\newcommand{\atDenison}{\affiliation{Dept.~of Physics and Astronomy, Denison University, Granville, Ohio 43023}}
\newcommand{\atNDL}{\affiliation{National Nano Device Laboratories, Hsinchu 300, Taiwan}}

\begin{document}

\title{Sensitivity of the As-Built Askaryan Radio Array to Ultra-High Energy Neutrinos}


 \author{N.~Alden}\atUC
 \author{S.~Ali}\atKU
 \author{P.~Allison}\atOSU
 \author{J.J.~Beatty}\atOSU
 \author{D.Z.~Besson}\atKU
 \author{A.~Bishop}\email{abigail.bishop@wisc.edu}\atUW
 \author{P.~Chen}\atNTU
 \author{Y.C.~Chen}\atNTU
 \author{Y.-C.~Chen}\atNTU
 \author{S.~Chiche}\atULB
 \author{B.A.~Clark}\atUMD
 \author{A.~Connolly}\atOSU
 \author{K.~Couberly}\atKU
 \author{L.~Cremonesi}\atUCL
 \author{A.~Cummings}\atPSUigc\atPSUphys\atPSUast
 \author{P.~Dasgupta}\atOSU
 \author{R.~Debolt}\atOSU
 \author{S.~de~Kockere}\atVUB
 \author{K.D.~de~Vries}\atVUB
 \author{C.~Deaconu}\atUC
 \author{M.A.~DuVernois}\atUW
 \author{J.~Flaherty}\atOSU
 \author{E.~Friedman}\atUMD
 \author{R.~Gaior}\atChiba
 \author{P.~Giri}\atUNL
 \author{J.~Hanson}\atWhittier
 \author{N.~Harty}\atUD
 \author{K.D.~Hoffman}\atUMD
 \author{M.-H.~Huang}\atNTU\atNUU
 \author{K.~Hughes}\atOSU
 \author{A.~Ishihara}\atChiba
 \author{A.~Karle}\atUW
 \author{J.L.~Kelley}\atUW
 \author{K.-C.~Kim}\atUMD
 \author{M.-C.~Kim}\atChiba
 \author{I.~Kravchenko}\atUNL
 \author{R.~Krebs}\email{ryan.j.krebs@psu.edu}\atPSUigc\atPSUphys
 \author{C.Y.~Kuo}\atNTU
 \author{U.A.~Latif}\atVUB
 \author{C.H.~Liu}\atUNL
 \author{T.C.~Liu}\atNTU\atNPU
 \author{W.~Luszczak}\atOSU
 \author{A.~Machtay}\atOSU
 \author{M.S.~Muzio}\email{muzio@wisc.edu}\atUW\atPSUigc\atPSUphys\atPSUast
 \author{J.~Nam}\atNTU
 \author{R.J.~Nichol}\atUCL
 \author{A.~Novikov}\atUD
 \author{A.~Nozdrina}\atOSU
 \author{E.~Oberla}\atUC
 \author{C.W.~Pai}\atNTU
 \author{Y.~Pan}\atUD
 \author{C.~Pfendner}\atDenison
 \author{N.~Punsuebsay}\atUD
 \author{J.~Roth}\atUD
 \author{A.~Salcedo-Gomez}\email{salcedogomez.1@osu.edu}\atOSU
 \author{D.~Seckel}\atUD
 \author{M.F.H.~Seikh}\atKU
 \author{Y.-S.~Shiao}\atNTU\atNDL
 \author{J.~Stethem}\atOSU
 \author{S.C.~Su}\atNTU
 \author{S.~Toscano}\atULB
 \author{J.~Torres}\atOSU
 \author{J.~Touart}\atUMD
 \author{N.~van~Eijndhoven}\atVUB
 \author{A.~Vieregg}\atUC
 \author{M.~Vilarino~Fostier}\atULB
 \author{M.-Z.~Wang}\atNTU
 \author{S.-H.~Wang}\atNTU
 \author{P.~Windischhofer}\atUC
 \author{S.A.~Wissel}\atPSUigc\atPSUphys\atPSUast
 \author{C.~Xie}\atUCL
 \author{S.~Yoshida}\atChiba
 \author{R.~Young}\atKU
\collaboration{ARA Collaboration}\noaffiliation

\date{\today}

\begin{abstract}
    The Askaryan Radio Array (ARA) is an ultra-high energy (UHE) neutrino observatory designed to detect the impulsive radio waves produced by relativistic particle cascades in the Antarctic glacial ice. Using a significantly enhanced simulation pipeline, which adds data-driven detector simulations and fully incorporates secondary particle production, we calculate the trigger-level acceptance of the entire array. We compare the resulting trigger-level sensitivity to constraints on the UHE neutrino flux from other detectors. Given its exposure from 2013 to 2023, we find that ARA achieves a world-leading sensitivity above about $10^{19}$~eV, depending on the details of the event selection used in a search. Moreover, we find that up to 13 neutrinos are predicted to have been observed in this period at trigger-level, assuming the most optimistic neutrino flux models. We show that observations of secondary particles account for up to 30\% of the total acceptance starting at $10^{19}$~eV, and we explore the potential signatures and implications of both multi-pulse (from direct and refracted pulses and/or from secondary particle interactions) and multi-station events. Finally, we comment on the implications of this study for the design of next-generation UHE neutrino experiments, in particular IceCube-Gen2 Radio.
\end{abstract}

\maketitle

\section{Introduction}

    Neutrinos are unique probes of the universe and at ultra-high energies (UHE; $E_{\nu}\ge 10^{17}$~eV) provide an unparalleled window into extreme astrophysical environments. Recently, the KM3NeT Collaboration observed the first UHE neutrino candidate with an energy of $220_{-110}^{+570}$~PeV~\cite{KM3NeT-neutrino}. Robust measurements of the UHE neutrino flux will enable these particles to play an important role in multimessenger astronomy, alongside cosmic rays, gamma rays, and gravitational waves. 
    
    A flux of UHE neutrinos is expected to be produced by ultra-high energy cosmic rays (UHECRs; $E_\mathrm{CR}\ge 10^{18}$~eV). Both photohadronic and hadronic interactions of UHECRs inside their sources can produce what are commonly referred to as astrophysical neutrinos. Similarly, photohadronic interactions with the cosmic microwave background~\cite{gzk-gk,gzk-z} and extragalactic background light during extragalactic propagation produce cosmogenic neutrinos. In this way, measurements or constraints on the UHE neutrino flux provide a direct probe into the UHECRs and properties of their sources. 
 
    Currently, the most stringent constraints on the flux of UHE neutrinos have been set by the Pierre Auger Observatory~\cite{pao-detector}, the IceCube Neutrino Observatory~\cite{icecube-ehe}, and the Antarctic Impulsive Transient Antenna (ANITA) I--IV~\cite{anita-iv} up to $10^{21}$~eV. To improve on these constraints, enormous $\mathcal{O}(100)~\text{km}^3$ detection volumes monitored for long periods of time are required. The Antarctic glacier is well suited for this purpose because the South Pole ice provides favorable $\mathcal{O}(1)~\text{km}$ attenuation lengths of radio waves, enabling efficient monitoring with radio antennas~\cite{SPice-attenuation}. These antennas can detect radio pulses produced by Askaryan radiation~\cite{Askaryan:1961pfb,Askaryan:1965}, the coherent pulse of radiation emitted along the Cherenkov cone by the cascade initiated by a particle interaction in a dielectric medium.
    The Askaryan Radio Array (ARA), an in-ice radio array at the South Pole operational since 2013, was designed to observe UHE neutrinos using this detection principle.
    
    In this paper, we present a detailed evaluation of ARA's sensitivity to UHE neutrinos based on the array's livetime from 2013 to 2023. This study provides a detailed understanding of the array's performance and serves as a foundation for a forthcoming diffuse neutrino flux search, which will be ARA's most sensitive neutrino search to date, using more than a decade of data~\cite{ara-5sa-icrc}.  
    Section~\ref{sec:background} provides a brief review of the relevant interaction processes by which UHE neutrinos in ice produce radio-frequency Askaryan radiation.
    In Section~\ref{sec:ara}, we describe the ARA detector itself.
    In Section~\ref{sec:simulation}, we explain the generation of simulated neutrinos, detector simulation, and calculation of ARA's acceptance in the data-taking period between 2013 and 2023. 
    In Section~\ref{sec:results}, we present ARA's trigger-level sensitivity and expected number of events predicted for several neutrino flux models.
    In Section~\ref{sec:discussion}, we discuss the signatures and implications of events with multiple observed pulses (due to propagation effects or observing several cascades from the same event) and multi-station detections, and how these results could influence designs for future detectors such as the IceCube-Gen2 radio array. 
    Finally, we give our conclusions in Section~\ref{sec:conclusion}. 
    In Appendix~\ref{app:comp}, we compare results of this work to previous ARA acceptance calculations, and in Appendix~\ref{app:zenith_acceptance}, we present the angular sensitivity of ARA.
    We note that Appendix~\ref{app:AraSim} provides a detailed discussion of \arasim{}, ARA's detector simulation software.

\section{In-Ice UHE Neutrino Interactions}\label{sec:background}

    Askaryan radiation is produced by particle showers initiated by relativistic charged particle interactions in dense dielectric media, where a negative charge excess builds as the shower evolves. This charge excess results in a coherent nanosecond-scale pulse of radio emission along the Cherenkov cone. 
    Askaryan emission was first observed in a beam test at SLAC National Accelerator Laboratory in 2001~\cite{slac} and recently in nature through observations of cosmic ray air showers penetrating the ice by ARA~\cite{ara-cosmic-ray-askaryan}. 
    
    Two distinct classes of particle cascades are produced by neutrino and secondary lepton interactions: hadronic showers, initiated by the hadrons generated by deep inelastic scattering and photonuclear interactions; and electromagnetic (EM) showers, initiated by electrons/positrons and photons. EM showers are less dense compared to hadronic showers, due to the smaller average multiplicity of their interactions and the Landau–Pomeranchuk–Migdal (LPM) effect~\cite{lpm:lp, lpm:m} at ultra-high energies. 
    The radio emission of the hadronic and EM showers must be modeled separately due to these differences in their shower profiles, as discussed in Ref.~\cite{Alvarez-Muniz:2020ary} and in Appendix~\ref{app:askaryan_model}.

    \begin{figure}
        \centering
        \includegraphics[width=\linewidth]{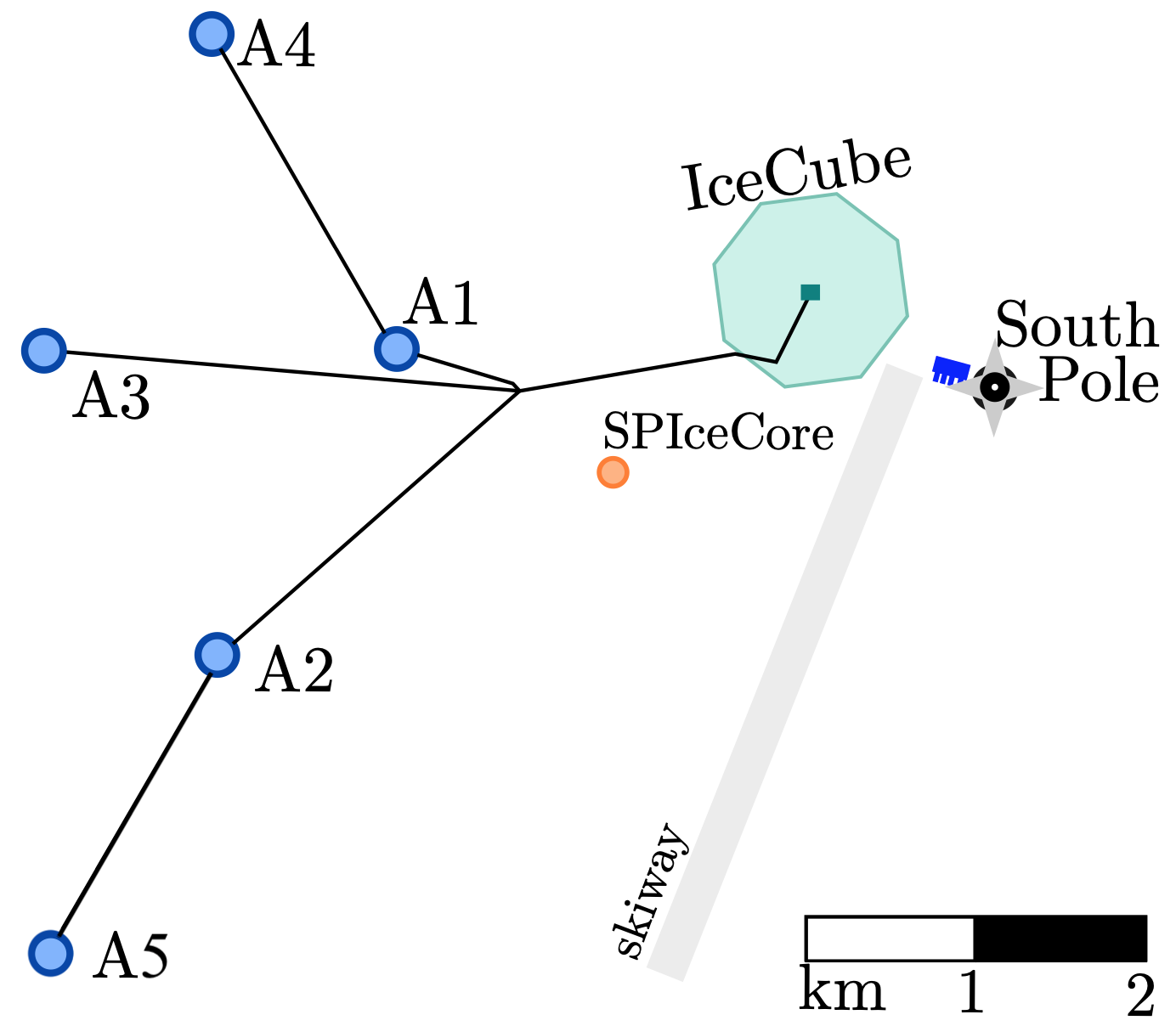}
        \caption{
            \label{fig:ARA_map}
            Map of the five ARA stations relative to the IceCube Neutrino Observatory and South Pole. 
        }
    \end{figure}
      
    At UHEs, neutrinos dominantly interact through deep inelastic scatterings via charged current (CC) and neutral current (NC) interactions. These represent the exchange of a charged $W$ boson or a neutral $Z$ boson with target nucleons in the medium, respectively. NC interactions initiate hadronic particle showers, generated from the recoiling nucleus, while the down-scattered neutrino continues to propagate. On the other hand, CC interactions produce an outgoing charged lepton ($e$, $\mu$, or $\tau$ as determined by the primary neutrino flavor), in addition to the hadronic shower. These outgoing leptons are referred to as secondary particles. 

    At UHEs, electrons do not need to be simulated separately and can be replaced by an EM shower, since they do not propagate a significant distance before initiating an EM shower. Outgoing $\mu$ and $\tau$ leptons, instead, must be simulated individually due to their long interaction and decay lengths. 
    These particles carry a fraction of the primary neutrino energy, determined by their inelasticity distribution, which is subsequently deposited into the ice through stochastic energy losses, including through bremsstrahlung, pair-production, and photonuclear processes. 
    Most stochastic losses do not deposit enough energy to generate a detectable impulsive signal, however catastrophic losses comparable to the charged lepton's energy can be observed by radio detectors~\cite{Cummings:2023iuw}. 
    In addition, the $\tau$ leptons produce another shower when they decay.

    Particle cascades initiated by secondary particle interactions and decays can significantly contribute to the sensitivity of neutrino detectors~\cite{Garcia-Fernandez:2020dhb,Coleman-flavor} and provide opportunities for flavor identification in radio detectors~\cite{Coleman-flavor}. Previous studies~\cite{Cummings:2023dbj, ARA:2023vaf} of ARA's total neutrino sensitivity, using an idealized detector simulation with the \texttt{PyREx} package,\footnote{\url{https://github.com/abigailbishop/pyrex}} showed that $25\%$ of observable cascades are due to secondary particle cascades when folded with the flux model of Ref.~\cite{flux-kotera}. These studies simulated cascades individually, did not incorporate detector deadtime, and did not consider the effects of observing multiple subsequent cascades from the same particle (such as multiple stochastic losses from a UHE muon) in a single waveform.
        
\section{The Askaryan Radio Array}\label{sec:ara}

    ARA is located near the geographic South Pole and consists of five autonomous stations (A1--A5). Figure~\ref{fig:ARA_map} shows the locations of each station in relation to the IceCube Neutrino Observatory. ARA stations are on a hexagonal grid with $2$~km spacing, chosen to maximize the sensitivity to $1$~EeV neutrinos~\cite{ara-design}. Power and communications are provided by cabling to the IceCube Laboratory, which also facilitates data storage and transfer.

    Each ARA station~\cite{ara-design} has four boreholes drilled to approximately $100$~m depth for A1 and $200$~m depth for A2--A5. Each borehole hosts two pairs of antennas deployed to $80$~m (A1) or $190$~m depth (A2--A5) and separated by 20--30~m vertically. These antenna pairs consist of one antenna sensitive to vertically polarized (VPol) signals and another to horizontally polarized (HPol) signals. The antenna pairs form an approximately rectangular lattice with $14$~m (A1--A3) or $30$~m (A4--A5) sides horizontally, which provides three-dimensional polarization and timing information for event reconstruction. The VPol antennas use a bicone design, while the HPol antennas employ a quad-slot design adapted to the 15~cm diameter borehole. These antennas are sensitive in the 150--850~MHz band, where Askaryan radiation is expected to be the strongest~\cite{ara-reco-prototypestation}. Each station triggers when three same-polarization channels detect excess power over a threshold within a configurable coincidence window of approximately 170~ns. This threshold is adjusted to maintain a $\sim6$~Hz trigger rate.

    \begin{figure}
        \centering
        \includegraphics[width=\linewidth]{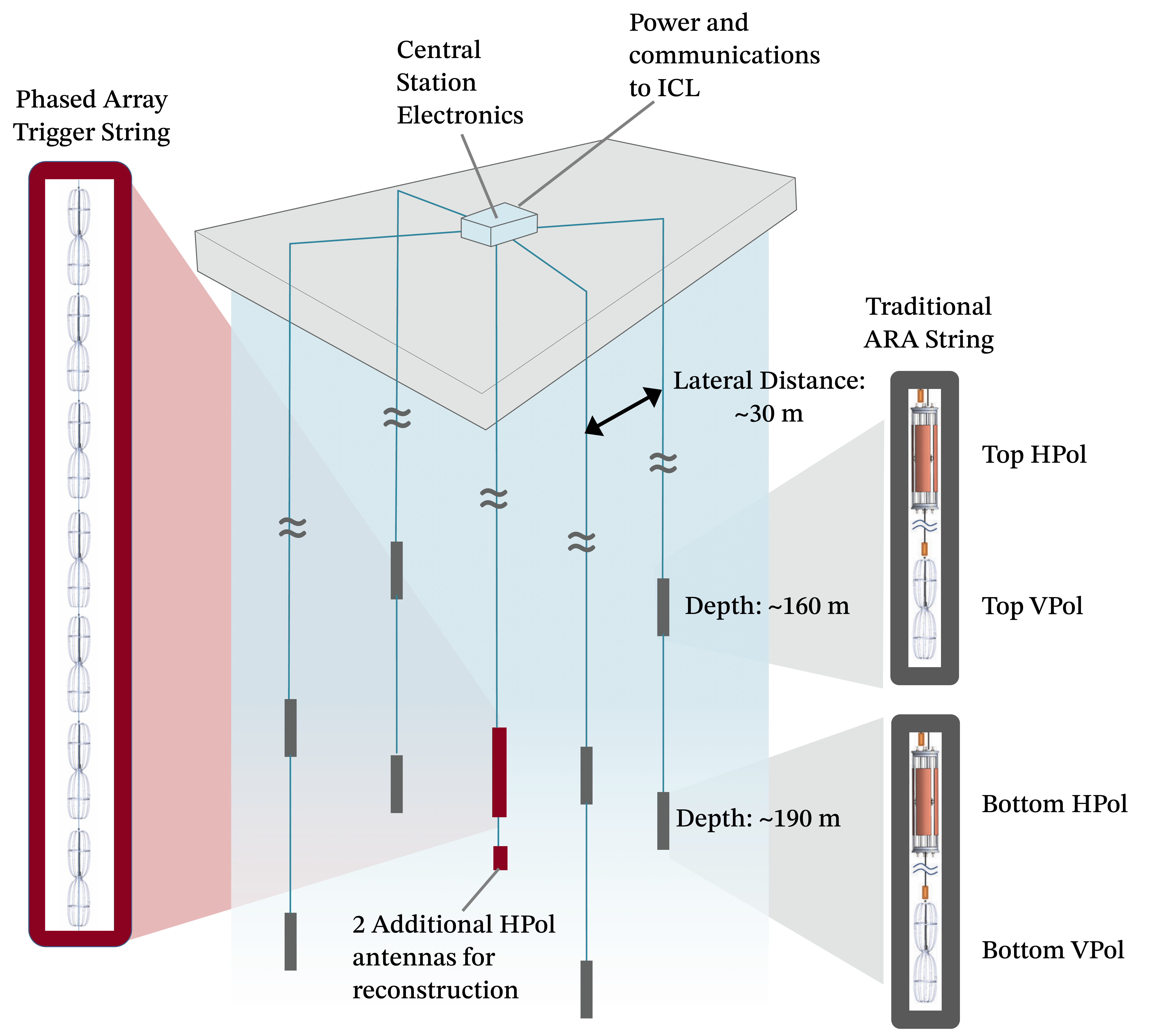}
        \caption{
            \label{fig:ARA_station}
            ARA's station layout for the example of A5. Traditional stations have 4 holes drilled up to 200~m into the ice and instrumented with radio antennas up to 190~m. A5 additionally has the phased array (PA) detector string deployed in the center of the 4 outer strings, also to a depth of 190~m.
            A1--A4 share this layout but do not have the central PA string. Additionally, A1 is deployed to a maximum depth of 80~m and A1--A3 have an installed lateral distance baseline of 14~m.
        }
    \end{figure}

    \begin{figure*}
        \centering
        \includegraphics[width=\linewidth]{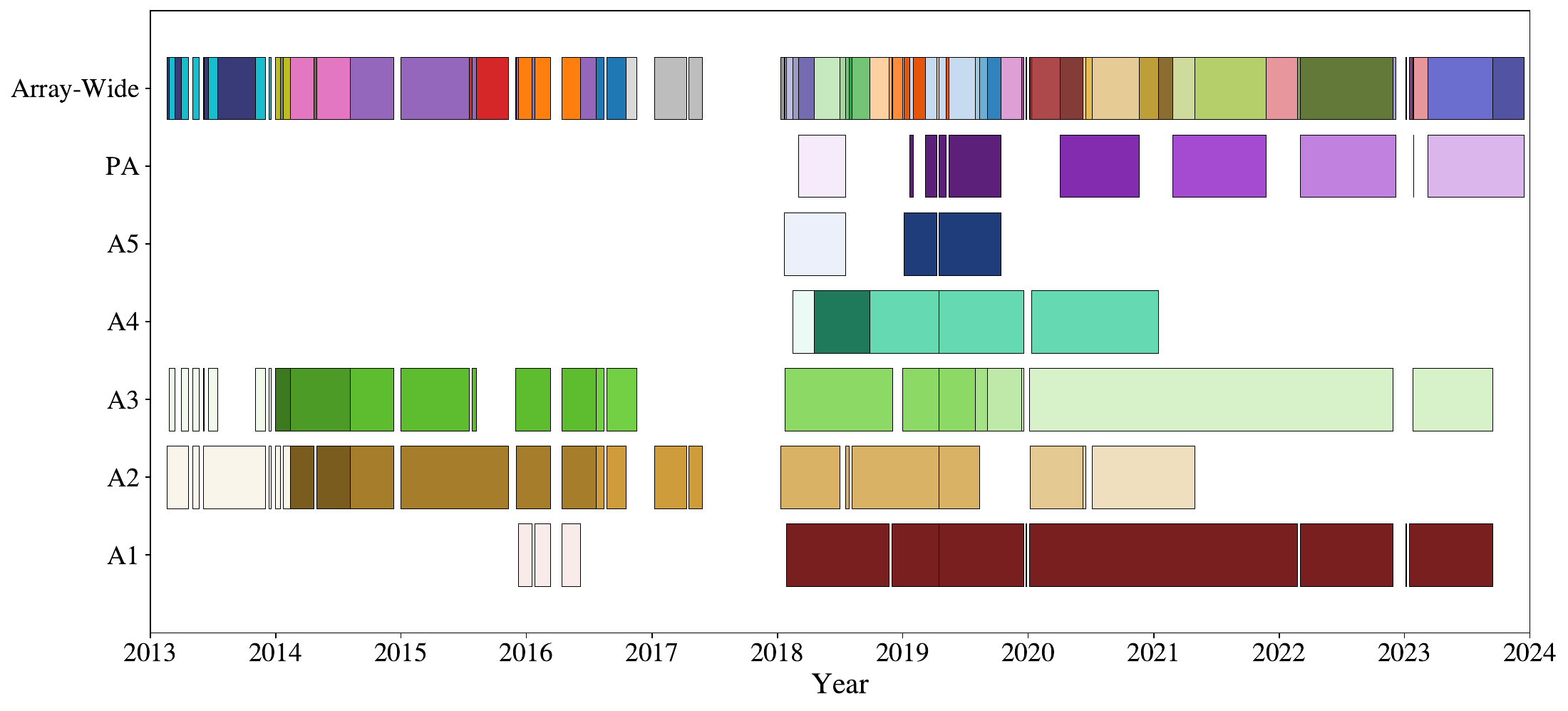}
        \caption{
            \label{fig:ARA_configurations}
            The configurations of ARA from 2013 to 2023. Station-level configurations for ARA stations A1--A5 and PA are shown in the bottom 6 rows as a function of calendar year. Combinations of these configurations result in $50$ distinct array-wide configurations shown in the topmost row, distinguished by color. Colors that repeat represent the same configuration.
        }
    \end{figure*}

    In addition to the traditional station design described above, A5 has an additional, central $200$~m borehole which hosts an independent phased array (PA) detector string. This string consists of seven VPol antennas and two HPol antennas, each separated by 1--2~m spacing. This densely instrumented string minimizes differences in signal between channels, allowing for their waveforms to be coherently summed to reduce the trigger threshold, enhancing its sensitivity~\cite{ara-pa-design}. The PA trigger is formed using only the VPol channels, and its threshold is adjusted to maintain a $\sim11$~Hz trigger rate. While A5 and PA are co-located, we refer to them throughout as separate detectors as they have independent data acquisition (DAQ) systems. Notably, one VPol channel from A5 was connected to the PA DAQ in 2018--2019, and an additional six were connected in 2019--2020. These channels provide improved event vertex reconstruction, particularly in azimuth. 
    
    In addition to these detection strings, each station includes in-ice calibration pulsers, deployed in their own boreholes, which are used to verify the station geometry and monitor each station's performance. Figure~\ref{fig:ARA_station} shows a diagram of the A5 station. A1--A4 are similar but do not have the central PA string and A4 only has $3$~detection strings.

    ARA has collected data since 2011, after deploying a prototype Testbed station in the 2010--2011 Antarctic summer season~\cite{ara-reco-prototypestation}. The main array was constructed over several deployment seasons: A1 in 2011--2012; A2 \& A3 in 2012--2013; and A4 \& A5 (including PA) in 2017--2018. 

    To correctly describe ARA's time-dependent sensitivity, each station's livetime is grouped into periods, called configurations. Each of these configurations is defined by determining periods when the station's hardware, detector settings, and overall noise environment were approximately stable. Each configuration is independently simulated using configuration-specific models of the detector settings, signal chain gain, and noise (see Section~\ref{sec:simulation}).

    These station-level configurations are then combined into array-wide configurations --- periods when the underlying number of stations taking data and their station-level configurations are stable. From 2013 to 2023, ARA had $50$ distinct array-wide configurations. Figure~\ref{fig:ARA_configurations} summarizes both the station-level and array-wide configurations in this period. The upcoming analysis of data taken between 2013 and 2023 is expected to be the most sensitive search for UHE neutrinos conducted by any in-ice radio detector to date~\cite{ara-5sa-icrc}.

\section{Simulation Setup}\label{sec:simulation}

    \par
    The neutrino simulations used to calculate ARA's sensitivity are divided into three parts: event generation, detector simulation, and sensitivity calculation. 
    First, \nuleptonsim{}~\cite{Cummings:2023iuw} performs a forward propagation of primary neutrinos through the Earth. This process generates lists of particle interactions cataloging their defining characteristics (such as energy, location, and direction) for UHE neutrinos and their outgoing particles. These lists of particle interactions, referred to as event libraries, form a common array-wide library of events.
    Next, these event libraries are passed to \arasim{} for signal generation and propagation, and for the detector and trigger simulation. 
    To reduce the overall computational expense, we perform single-station simulations of the array-wide event libraries for every station in each of their stable configurations. 
    Finally, ARA's acceptance, sensitivity, and estimated number of events are calculated by combining the triggers from station-level simulations into a unified trigger set for each array-wide configuration. 
    Each of these parts is discussed in detail in the following subsections. 
    
	\subsection{Event Generation with \nuleptonsim{} \label{sec:NLS}}
    
        Events are produced via a rejection sampling using the \nuleptonsim{} Monte Carlo simulation code. This code generates and forward propagates neutrinos and secondary particles through the Earth at the individual interaction level. This process is broken into two main steps.

\begin{figure}
            \includegraphics[width=\linewidth]{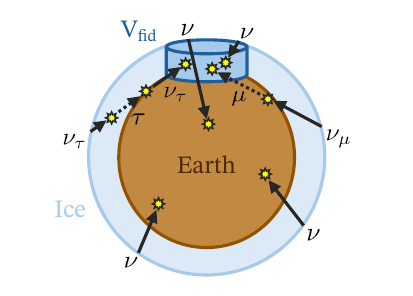}
            \caption{
                \label{fig:event_gen_diagram}
                Diagram of the \nuleptonsim{} event generation geometry. Neutrinos are thrown towards points uniformly sampled in the fiducial volume $V_{\text{fid}}$. 
                The circular face of $V_{\text{fid}}$ is centered on A2 (see the ARA station layout in Fig.~\ref{fig:ARA_map}). 
                This cartoon shows examples of neutrinos and secondaries interacting with the Earth and ice, including an example of tau neutrino regeneration (upper left) and an outgoing muon (upper right). 
                The yellow stars represent particle interactions. 
                Only the interactions that fall within $V_{\text{fid}}$ are saved to the event library and passed to the detector simulation.
            }
        \end{figure}

        First, a point is uniformly sampled in a fiducial cylinder --- the top face of which is centered on the centrally located A2 station. The size of the cylinder depends on the primary neutrino's energy,\footnote{
            We approximate stations as independent (i.e.\ they have zero overlap in their triggered neutrinos) for primary neutrino energies $\leq 10^{17.0}$~eV. We therefore use smaller fiducial cylinders (centered on each station) at these energies to improve the computational efficiency of simulations. Specifically, we use the following cylinder parameters for simulations of primary neutrinos with discrete energies: $E_\nu = 10^{16}$~eV, $r=1.5$~km, $h=0.75$~km; $E_\nu = 10^{16.5}$~eV, $r=3.0$~km, $h=1.5$~km; and $E_\nu = 10^{17}$~eV, $r=5.0$~km, $h=2.0$~km. These parameters were chosen to minimize the size of the fiducial volume while still fully encompassing each station's detection volume.
            \label{fn:lowEsims}
        } but for energies greater than $10^{17}$~eV, the cylinder has a $15$~km radius and $3$~km height. 
        A chord is then drawn from this point in a direction sampled uniformly in azimuth and in cosine zenith. This chord is extended until it intersects the Earth's surface. The primary neutrino is thrown towards the detector along this chord, starting from the intersection point with Earth's surface. This process guarantees that only trajectories passing through the fiducial volume are sampled. A diagram of the geometry used is shown in Fig.~\ref{fig:event_gen_diagram}.

        Second, the neutrino is forward propagated towards the fiducial volume from the initial Earth intersection along the chosen chord. 
        The Earth is modeled using a layer-averaged density profile derived from PREM~\cite{dziewonski-anderson-1981} to allow for rapid calculation of propagation lengths. 
        To model the Antarctic glacial ice, the Earth is enveloped by a 3~km thick shell of ice with a uniform density of $0.92$~g/cm$^3$.
        Individual neutrino interaction distances and interaction types are sampled according to the neutrino-nucleon cross section of Ref.~\cite{ctw}, while cascade and secondary particle energies are determined using inelasticity tables from the CTEQ5 parton distribution~\cite{cteq5}. Secondary particle interactions are calculated using parameterized differential cross sections.
        Particle propagation stops when particles exit the fiducial volume or when their energy drops below $E_\mathrm{th} = \min(10~\text{PeV}, E_\nu/10)$. This energy threshold was chosen to minimize computation time while retaining all particles that contribute to ARA's sensitivity.
        
        Each primary neutrino is assigned a unique tracking ID that is inherited by all particles resulting from interactions that the primary neutrino and its secondary particles undergo. Any interaction that occurs in the fiducial volume with $E \geq E_\mathrm{th}$ is recorded by saving: the primary neutrino's tracking ID, flavor, and energy along with interaction-specific information, including the initiating particle flavor \& propagation direction, interaction vertex location, interaction type \& inelasticity, and shower type. This information for each interaction in the fiducial volume constitutes the event library passed to \arasim{}. Event libraries are generated for both the primary-only case (i.e., considering only neutrinos whose first interaction is in the fiducial volume and ignoring any further interactions of the neutrino or its secondaries) and the complete primary-and-secondary case to study the effects of including secondary particle interactions.

        We note that the procedure for event library generation described above amounts to a rejection sampling procedure that generates a sample of interaction vertex sets (which share a common primary neutrino) in the fiducial volume. This space is most straightforwardly sampled via a forward-propagation Monte Carlo since it depends upon the full history of interactions up to and within the fiducial volume. Only primary neutrinos that result in interaction vertices within this fiducial volume are represented in the (biased) sample simulated by \arasim{}, while all other primary neutrinos are ``rejected'' from this sample. This bias is later corrected via Monte Carlo weighting.
         
        In addition to event library generation, \nuleptonsim{} is also used to calculate the solid angle-averaged probability, $\langle p_\mathrm{acc}\rangle$, for a primary neutrino to produce at least one interaction in the fiducial volume. This is calculated via dedicated auxiliary simulations for each neutrino flavor, charge, and energy.  
        
	\subsection{Detector Response Simulation with \arasim{} \label{sec:AraSim}}

        The library of neutrino interactions and secondary particle interactions, generated in the previous step via \nuleptonsim{}, is then provided to \arasim{}, which simulates the radio emission, generated by the cascades from these interactions, and the detector response. 
        \arasim{} has been significantly updated for this study, compared to the version used in Refs.~\cite{ara-a23, ara-pa}. These updates include the incorporation of data-driven detector modeling and the realistic simulation of multi-pulse signals in the detector. 
        Here, we summarize the important updates made for this work. A detailed description of \arasim{} is presented in Appendix~\ref{app:AraSim}.
        
        \begin{figure}
            \centering
            \includegraphics[width=\linewidth]{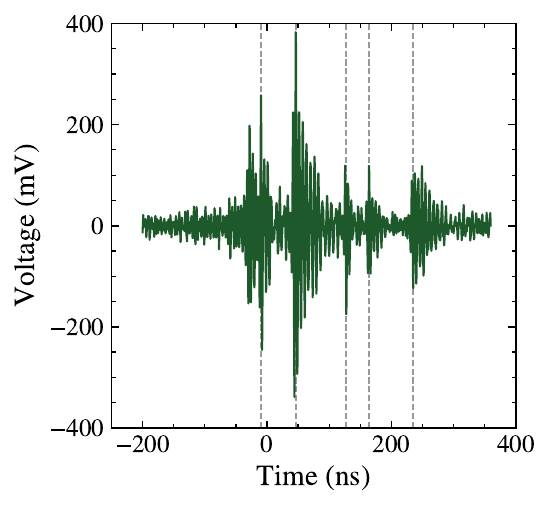}
            \caption{ 
                \label{fig:waveform}
                Simulated voltage trace of one channel on station A3 containing multiple pulses from primary and secondary interactions.
                The primary neutrino which generated this event was a $10^{19}$~eV muon neutrino.
                Distinct pulses are indicated by vertical dashed lines. Pulses were found according to the procedure described in Section~\ref{sec:discussion}. The two pulses nearest to the first vertical dashed line are indistinguishable under this procedure.
            }
        \end{figure}

        For each interaction in the event library, \arasim{} reads in the location, direction, energy, interaction type, and inelasticity of the interaction, as well as the primary neutrino tracking number. From this data, the electric field produced by the cascade is generated, and ray paths connecting the interaction to each antenna of the station are calculated. This electric field is attenuated while traveling along each ray path due to the dielectric properties of the ice.
        
        We convert from the electric field to the expected voltage by applying the antenna's effective height. 
        Voltage traces from each interaction associated with the same primary neutrino (i.e., voltages from all interactions with the same tracking ID) are added together to form one combined trace per antenna for that event, with relative delays between signals calculated from their interaction times and travel times from ray tracing.
        The combined traces then pass through the station electronics chain, and simulated system-amplified thermal noise is added to create waveforms for the whole event.
        These combined traces are as long as necessary to encompass all cascades from the event.
        
        Finally, a simulation of the trigger is successively applied to the trace. 
        For stations A1--A5, a trigger is formed when at least three antennas of the same polarization detect a signal-to-noise ratio\footnote{We define the SNR as half the peak-to-peak voltage in a small time window divided by the root mean square voltage outside the signal region.} (SNR) greater than $\sim6$ within a coincidence window, which varies station to station. 
        For the PA, a potential trigger's outcome is determined by accepting the trigger with a probability equal to the detector's measured trigger efficiency~\cite{ara-pa-design} for the signal-only SNR of the top-most VPol antenna.

        If a station triggers, \arasim{} saves the event information and the waveforms recorded by each antenna within a 400--600~ns readout window (depending on the station and configuration) around the trigger position. After accounting for the detector's deadtime, the trigger simulation continues until the full signal waveform has been analyzed. An example of a triggered waveform in one antenna is shown in Fig.~\ref{fig:waveform} for station A3. This waveform contains pulses from multiple interactions generated by a primary muon neutrino event and its secondary particles.

        \begin{figure}
            \centering
            \includegraphics[width=\linewidth]{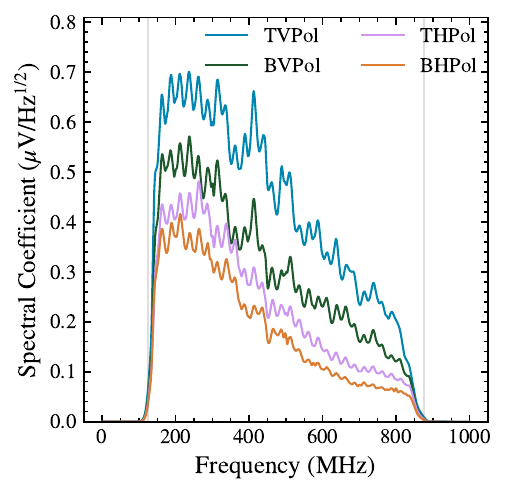}
            \caption{
                \label{fig:noise_model}
                Comparison of noise models in \arasim{}. The spectral coefficient (i.e.\ the Rayleigh distribution scale parameter for a frequency bin) is shown as a function of frequency for the data-driven models for the example of String 1 of A3 in configuration 1. Data-driven models are presented for each channel: top VPol (TVPol), bottom VPol (BVPol), top HPol (THPol), and bottom HPol (BHPol). Models are shown after the system response has been applied. A detailed description of these models can be found in Appendix~\ref{app:detector_modeling}.}
        \end{figure}
        
        \begin{figure}
            \centering
            \includegraphics[width=\linewidth]{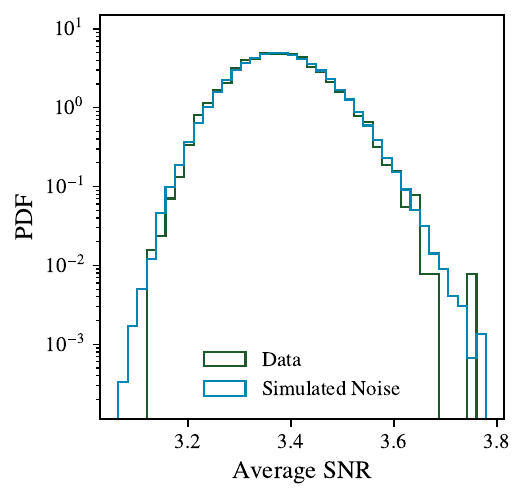}
            \caption{
                \label{fig:noise_comp}
                Signal-to-noise ratio (SNR) averaged over all VPol channels for simulated noise (blue) and real noise-like (green) events in configuration 4 of the Phased Array detector.}
        \end{figure}

        Recent updates to the simulation framework have significantly improved the fidelity of the detector modeling by incorporating refined antenna gain, system gain, and noise models. The antenna gain models used in simulation are now derived from in-air anechoic chamber measurements and then transformed to their corresponding in-ice responses (see Appendix~\ref{app:antenna}), rather than from in-ice antenna simulations as used previously. These measurements were taken for three antenna types: HPols, bottom VPols, and top VPols (distinguished because additional cables passing through the top VPols affect their beam pattern). 
        
        Additionally, \arasim{} now uses data-driven system gain and noise models based on in-situ measurements, rather than previously used models derived from \texttt{Qucs}\footnote{https://github.com/Qucs/qucs} electronics simulations and theoretical Johnson-Nyquist noise models, respectively. These system gain and noise models are constructed for each antenna, station, and detector configuration, enabling realistic simulations of the configuration-dependent detector response across the full data-taking period. Figure~\ref{fig:noise_model} illustrates the new data-driven channel-dependent noise models. A similar comparison of gain models can be found in Appendix~\ref{app:detector_modeling}.

        The agreement between data and simulations enabled by these updates is illustrated in Fig.~\ref{fig:noise_comp}, in terms of the average SNR of waveforms in VPol channels for simulated thermal noise events and thermal noise-like data.
        A detailed discussion of these updates can be found in Appendix~\ref{app:AraSim}.
        
    \subsection{Calculation of Detector Sensitivity \label{sec:calculations}}

        As described in the previous sections, for computational efficiency, each station and configuration is simulated separately using a fixed event library, rather than simulating all $50$~array-wide configurations directly. Array-wide quantities are calculated via a dedicated processing pipeline that combines station-level triggers into a unified list of ``array-level” triggers for each configuration of the array. This pipeline ensures that primary neutrinos that lead to multiple triggers across the array are counted only once. This procedure provides a flexible framework for calculating the array-wide acceptance from the responses of the individual stations in each of their configurations.

        \par   
        We calculate the array-wide acceptance from a sample of primary neutrinos and their secondary particles generating at least one interaction inside the fiducial volume. This event sample is generated via a rejection sampling using \nuleptonsim{}, as described in Sec.~\ref{sec:NLS}, and then simulated through  \arasim{} to determine the trigger outcome of each event, as described in Sec.~\ref{sec:AraSim}. Within this framework, the Monte Carlo estimator for the acceptance is given by

        \begin{align}\label{eq:acceptance}
            \mathcal{A}(E) = \frac{1}{N} \sum_i^{N_\mathrm{trig}} w_i~,
        \end{align}
        where $N$ is the number of unique primary neutrino events in the event library passed to \arasim{} (N.B.\ this is not the same as the total number of primary neutrinos simulated by \nuleptonsim{}), $N_\mathrm{trig}$ is the number of triggering primary neutrino interactions, and $w_i$ is the weight of the $i$th triggered event.\footnote{At $E\leq 10^{17}$~eV, stations are approximated as independent. Therefore, their acceptances are calculated individually and then summed to find the array-wide acceptance.} The weights for events are calculated as
        
        \begin{align}\label{eq:weights}
            w_i = \frac{4\pi V_\mathrm{fid} \langle p_\mathrm{acc} \rangle}{L_i}~,
        \end{align}
        where $V_\mathrm{fid}$ is the fiducial volume, $\langle p_\mathrm{acc} \rangle$ is the solid angle-averaged probability for a primary to produce an interaction vertex in the fiducial volume (i.e. to be accepted in the rejection sampling scheme), and $L_i$ is the length of the chord through the fiducial volume along which particles from the $i$th primary neutrino propagate. Consequently, $w_i$ has units of area times solid angle. The numerator in Eq.~\eqref{eq:weights} corresponds to the normalization factor of the sampling distribution passed to \arasim{}, while the denominator corrects for the over-representation of primary particle interactions along chords that have longer path lengths through the fiducial volume. 
        
        \par
        To determine the trigger-level sensitivity of the array, we first determine its total array-wide exposure. In the following discussion, we assume a flavor ratio of $\nu_e:\nu_\mu:\nu_\tau = 1:1:1$ and a neutrino-to-antineutrino ratio of $\nu : \bar{\nu} = 1:1$. Because the array operated in multiple configurations over time with changing trigger thresholds and livetimes, the acceptance in Eq.~\eqref{eq:acceptance} is generally time-dependent. We therefore define the trigger-level, flavor-averaged (over all 6 types of neutrinos) exposure as
        \begin{align}\label{eq:exposure_trigger}
            \mathcal{E}(E) = \sum_c \mathcal{A}_c(E) T_c~,
        \end{align}
        where the sum is over all array configurations and $\mathcal{A}_c$ and $T_c$ are the flavor-averaged acceptance and livetime of configuration $c$, respectively. To obtain the analysis-level exposure, we multiply each term in the sum in Eq.~\eqref{eq:exposure_trigger} by its corresponding signal efficiency, $\varepsilon_c(E)$ --- the fraction of triggered neutrinos preserved by a search's event selection.

        Given the total flavor-averaged exposure in Eq.~\eqref{eq:exposure_trigger}, we can construct an upper limit on the differential flux of diffuse neutrinos. We evaluate the all-flavor diffuse neutrino flux upper limit in decade-wide bins, assuming an $E^{-1}$ spectrum within the bin as
        \begin{align}
            \phi^\mathrm{UL}_{\nu+\bar{\nu}}(E_0) = \frac{N_\mathrm{UL}}{\ln(10) E_0 \mathcal{E}(E_0)}~,
        \end{align}
        where $N_\mathrm{UL}=2.44$ is the background-free $90\%$~confidence level (CL) upper limit on the true average number of neutrinos for zero observed events under the Feldman-Cousins construction~\cite{feldman-cousins}, and $E_0$ is the logarithmic center of the constrained energy bin. This trigger-level differential upper limit represents the maximum achievable sensitivity of the detector. 

        To mimic the sensitivity of a simplified neutrino search, we perform a toy event selection on the simulated neutrino events to determine the analysis-level exposure.  
        Events are selected as a signal if their SNR is incompatible with the thermal background and their reconstructed elevation angle is below the value where anthropogenic background signals are expected. Specifically, the following conditions are required: 
        \begin{itemize}
           \item The average SNR across antennas of the same polarization is required to be $\geq5$. 
           \item The interferometrically reconstructed~\cite{ara-reco-prototypestation} elevation angle is at least $5^\circ$ below the total internal reflection angle for the station,\footnote{
            Events are reconstructed to a hypothesized vertex position on a $300$~m radius sphere centered on the station's average antenna position. The event selection is applied to the straight-line (elevation) angle from the sphere's center to the reconstructed vertex position. The TIR angle reported in the text is the receipt angle at the sphere's center, which is then transformed into a reconstructed vertex position via ray tracing from the center to its edge. The straight-line angle to the ray-traced intersection is then used as the selection threshold.
           } corresponding to roughly $\theta \leq \theta_\mathrm{TIR} - 5^\circ \simeq 50^\circ$. The $5^\circ$ buffer is added to account for errors in angular reconstruction. 
           We expect that this is the deepest angle that horizontal rays from the surface could arrive from and thus drives this cut.
        \end{itemize}
    
        \noindent In Section~\ref{sec:results}, we compare the trigger-level sensitivity to the toy event selection's analysis-level sensitivity. A forthcoming publication will present the analysis-level sensitivity of a fully developed neutrino search. Primarily due to the high levels of anthropogenic backgrounds and incompletely characterized cosmic ray background, this forthcoming analysis uses a more conservative cut in the elevation angle that results in a reduced analysis-level sensitivity compared to the toy case presented here.
        Even so, real event selections used in previously published ARA searches~\cite{ara-a23,ara-pa} achieved up to 88\% of this simplified event selection's signal efficiency for $10^{21}$~eV neutrinos.\footnote{
            The toy analysis has an array-wide efficiency of 67\% for detecting $10^{21}$~eV neutrinos, while previous analyses of the PA and A2 had efficiencies of 91\%~\cite{ara-pa} and 54\%~\cite{ara-a23} to these neutrinos, respectively. In the 2013--2023 dataset, the PA had a livetime of 1421 days, and the total livetime of stations A1--A5 was 8878 days. Applying the A2 efficiency to A1--A5 and taking an average weighted by the livetime of each station, the estimated array-wide efficiency from previous results is 59\%.
        }

        \par
        Finally, the expected number of triggered events for a flux model is given by

        \begin{align}
            N = \int \phi(E) \mathcal{E}(E)~dE~,
        \end{align}
        where $\phi(E)$ is the total all-flavor neutrino flux model and $\mathcal{E}$ is the flavor-averaged exposure from Eq.~\eqref{eq:exposure_trigger}. 

        \begin{figure}
            \centering
            \includegraphics[width=\linewidth]{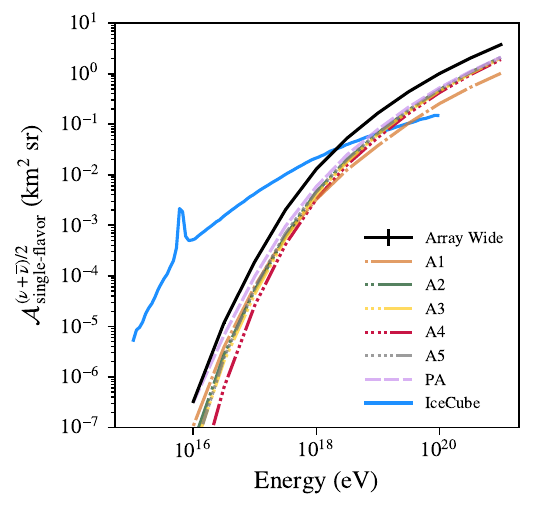}
            \caption{
                \label{fig:veffperstation}
                The trigger-level acceptance averaged over all 6 neutrino types and averaged over the livetime of the full array (black) compared to the analysis-level IceCube acceptance (blue)~\cite{icecube-ehe}.
                The trigger-level acceptance of each station (averaged over the livetime of each configuration) is also shown. 
            }
        \end{figure}
    
\section{Results}\label{sec:results}
    
        ARA's acceptance depends strongly on the properties of the individual stations, like their trigger type, depth, and number of strings. This is illustrated in  Fig.~\ref{fig:veffperstation}, where we show the flavor-averaged acceptance of each ARA station. 
        The PA has the largest acceptance at low energies due to its improved sensitivity to low SNR events. A1 has the smallest acceptance at high energies due to its shallow 80~m depth.
        A4 has the smallest acceptance at low energies due to its higher trigger threshold, caused by having only 3 strings to form triggers.
        Comparison of the A2 and PA acceptances to previously published results are given in Appendix~\ref{app:comp}.
        
        The array-wide acceptance is computed by averaging over all 50 array-wide configurations, weighted by the livetime of each configuration. 
        We also compute the analysis-level effective area as a function of neutrino arrival direction as a demonstration of ARA's angular sensitivity in Appendix~\ref{app:zenith_acceptance}.
    
\begin{figure}
            \centering
            \includegraphics[width=\linewidth]{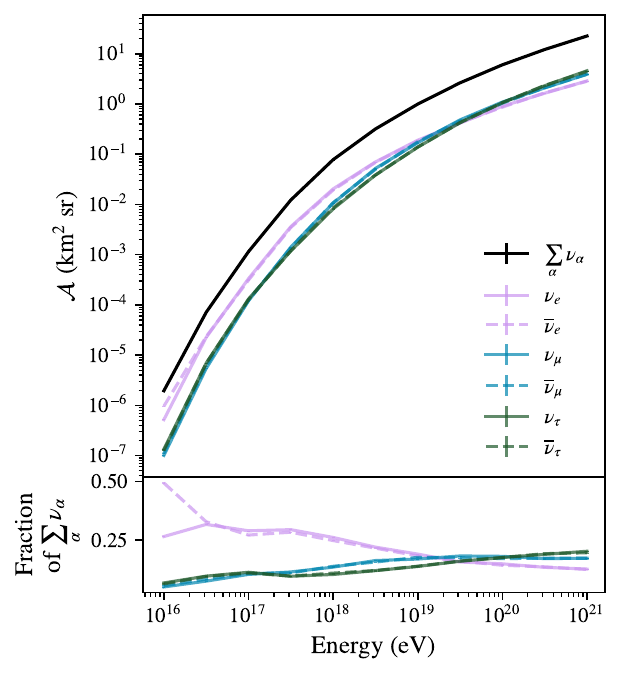}
            \caption{
                \label{fig:veffbyflavor}
                Top panel: trigger-level acceptance of ARA to neutrinos (solid lines) and anti-neutrinos (dashed lines). The acceptance is shown for electron neutrinos (purple), muon neutrinos (blue), and tau neutrinos (green) separately and summed together (black).
                Bottom panel: the fraction of each neutrino flavor relative to the sum.
            }
        \end{figure}

        ARA's acceptance to each neutrino flavor is shown in Fig.~\ref{fig:veffbyflavor}.
        Electron neutrinos dominate the acceptance up to approximately $10^{19.5}$~eV, because in CC interactions all of the energy from the neutrino is immediately deposited into the ice as a simultaneous hadronic and electromagnetic shower.
        Above $10^{19.5}$~eV, muon and tau neutrinos dominate the acceptance. In this regime, stochastic energy losses from muons and taus, as well as tau decays, produce secondary cascades that are sufficiently energetic to be observed by ARA, increasing their probability of detection.
        
\begin{figure}
            \centering
            \includegraphics[width=\linewidth]{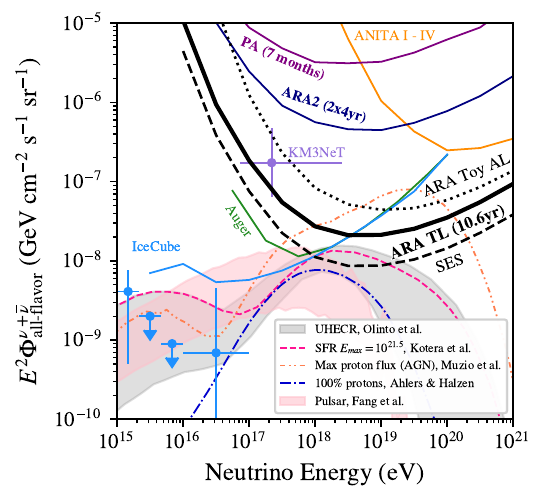}
            \caption{
                \label{fig:sensitivity}
                Array-wide sensitivity of ARA as characterized by $90\%$ CL upper limits. Upper limits for ARA are presented at both trigger-level (TL, solid black) and analysis-level (AL, dotted black), corresponding to the toy event selection using SNR $\geq 5$ and elevation $5^\circ$ below the total internal reflection angle. ARA's trigger-level single event sensitivity (SES, dashed black) is also shown. The trigger-level curves indicate the sensitivity accessible with a perfectly efficient analysis.
                Previous results from IceCube~\cite{icecube-ehe, icecube-combined-fit}, KM3NeT~\cite{KM3NeT-neutrino}, Auger~\cite{flux-auger}, ANITA I--IV~\cite{anita-iv}, and earlier ARA analyses (ARA2~\cite{ara-a23} and PA~\cite{ara-pa}) are also shown (colored solid).
                Several cosmogenic and astrophysical models~\cite{flux-kotera,flux-muzio,flux-ahlers-halzen,flux-fang-pulsars, Olinto:2011ng} of the UHE neutrino flux are presented. 
            }
        \end{figure}

        Figure~\ref{fig:sensitivity} shows the array-wide ARA sensitivity achieved between 2013 and 2023 characterized in terms of $90\%$ CL upper limits. Upper-limits are shown at trigger-level and at analysis-level for the toy event selection described in Section~\ref{sec:calculations}. 
        From this, one can see that ARA is the most sensitive UHE neutrino detector above about $10^{19}$~eV. 
        The expected number of events for several flux models~\cite{flux-kotera,flux-muzio,flux-ahlers-halzen,flux-fang-pulsars} included in this plot are listed in Table~\ref{tab:eventcounts}. This table includes expectations at both trigger- and analysis-level, as well as expectations for the number of events triggering more than one station.
        
        \begin{table}
            \centering
            \begin{tabularx}{\columnwidth}{>{\centering\arraybackslash}X|S[table-format=2.2]|S[table-format=1.2]|S[table-format=1.2]}
            \toprule
            \textbf{Flux Model} & \textbf{Trigger} & \textbf{Analysis} & \textbf{Multi-station}  \\
            \midrule
            Muzio~\cite{flux-muzio} & 13.01 & 6.28 & 1.82 \\
            Kotera~\cite{flux-kotera} & 2.27 & 0.95 & 0.20 \\
            100\% Protons~\cite{flux-ahlers-halzen} & 0.96 & 0.36 & 0.06 \\
            Fang Pulsars~\cite{flux-fang-pulsars} & 1.46 & 0.48 & 0.06 \\
            KM3NeT~\cite{KM3NeT-neutrino} & 13.71 & 4.24 & 0.36 \\
            \bottomrule
            \end{tabularx}
            \caption{
                \label{tab:eventcounts}
                Expected number of events for several UHE neutrino flux models, shown at both trigger-level and toy analysis-level. 
                The number of events at trigger-level predicted to be detected by two or more stations is also presented. 
                The expected number of events for the KM3NeT flux is calculated assuming an $E^{-2}$ spectrum over the energy range encompassed by its error bars.
            }
        \end{table}

        \begin{figure}
            \centering
            \includegraphics[width=\linewidth]{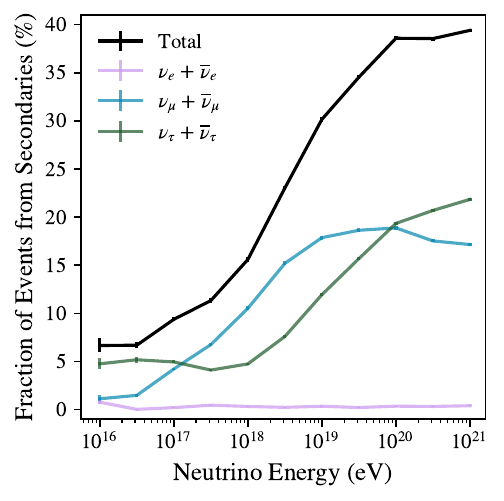}
            \caption{
                \label{fig:secondaries_contribution}
                Fraction of events contributed by secondary particle interactions at trigger-level. The contribution to the total acceptance (black) is broken down into contributions from electron neutrino (purple), muon neutrino (blue), and tau neutrino (green) secondaries. 
            }
        \end{figure}
        Figure~\ref{fig:secondaries_contribution} shows the percentage of the array-wide acceptance contributed by secondary particle interactions. We find that the contribution of secondary interactions increases with energy, with secondaries accounting for $40\%$ of ARA's total acceptance at $10^{21}$~eV. The contribution of secondaries by flavor is also shown in Fig.~\ref{fig:secondaries_contribution}. We see that electron neutrinos produce few secondary particles that contribute to ARA's acceptance, while muon and tau neutrino secondaries add significantly to the detector's overall sensitivity. 

        Figure~\ref{fig:multistation} shows the fraction of events that trigger two or more stations across the array. The share of events that generate multistation triggers increases with energy, reaching more than $35\%$ of all triggers at $10^{21}$~eV. This underscores that multistation triggers are not uncommon at high energies.

\section{Discussion}\label{sec:discussion}

        \par
        Beyond secondary interactions' significant contribution to ARA's overall sensitivity (see Fig.~\ref{fig:secondaries_contribution}), the results presented in Section~\ref{sec:results} highlight the unique reconstruction opportunities and challenges that they present, both at the individual-station and array-wide levels.

        \par
\begin{figure}
            \centering
            \includegraphics[width=\linewidth]{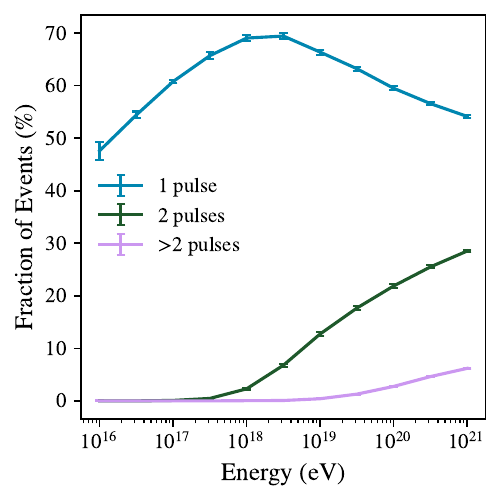}
            \caption{
                \label{fig:pulses_percentage}
                The fraction of events at trigger-level with one pulse (blue), two pulses (green), or greater than two pulses (purple) in their waveforms, for the example of A3 in configuration 6. One or two pulses in a waveform are possible from a single interaction, but more than two require pulses from at least two interactions to be present in the waveform. Details of the estimation can be found in the text. Note that the values of the lines at a particular energy need not sum to $100\%$, since pulses must have $\mathrm{SNR} \geq 5$ to be included in this estimation.
            }
        \end{figure}

        For favorable geometries, a series of primary and secondary interactions can generate multiple pulses in a single event readout. Each interaction potentially contributes up to two pulses, one from the ``direct'' ray path and one from the ``refracted'' ray path. These events can produce complicated voltage traces, such as the one shown in Fig.~\ref{fig:waveform}, which contains $5$~pulses from primary and secondary interactions. To understand their prevalence, we quantify the percentage of events that are multi-pulse by scanning all the channel waveforms of each event for pulses with $\mathrm{SNR} \geq 5$. The root mean square (RMS) voltage of each channel was taken from pure thermal noise simulations for this purpose, since multi-pulse traces can bias the estimation of the noise RMS. Additionally, an isolation condition is enforced by combining above-threshold pulse windows separated by less than $40$~ns, measured from the end of one window to the beginning of the next, to account for post-pulse ring-down structure. After merging, the peak position of a pulse is defined as the location of the maximum waveform amplitude within the merged window. The final number of pulses for the event is estimated from the channel with the maximum pulse count. The resulting number of multi-pulse events is shown in Fig.~\ref{fig:pulses_percentage} as a function of primary neutrino energy. In particular, some events are estimated to have zero pulses, due to the SNR threshold used.

        \par
        As can be seen from Fig.~\ref{fig:pulses_percentage}, multi-pulse events are not uncommon, making up $>30\%$ of events at $10^{21}$~eV. These multi-pulse events can provide critical information for reconstructing an event's arrival direction, energy, and flavor, as explored in previous studies~\cite{nuradiomc-glaserstudies, nuradiomc-secondaries}. However, such event morphologies also present challenges for neutrino searches. Vertex reconstruction, a critical step in typical neutrino searches, can be challenging for multi-pulse traces. Direct application of standard interferometric techniques to these events often results in poor-quality reconstructions, while more advanced techniques are technically complicated and computationally expensive. For this reason, neutrino searches must develop analysis tools with this class of events in mind to preserve the gains in sensitivity that they provide, particularly at high energies. 

        \begin{figure}
            \centering
            \includegraphics[width=\linewidth]{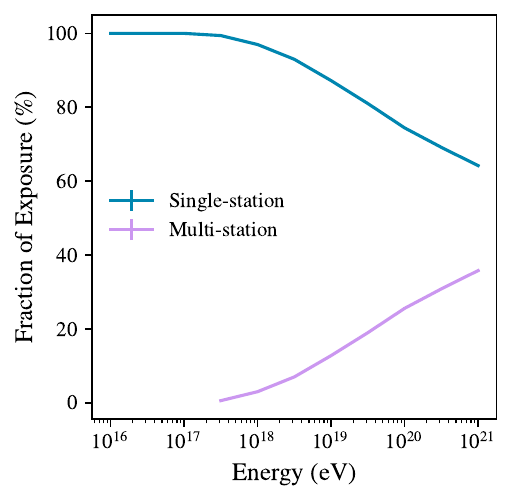}
            \caption{
                \label{fig:multistation}
                Fraction of events at trigger-level that generate triggers in only one station (blue) and in two or more stations (purple). 
                For neutrinos with energies $10^{16}-10^{17}$~eV, a multi-station trigger could not be determined as simulations were not array-wide.\footref{fn:lowEsims}
            }
        \end{figure}
        
        \par
\begin{figure}
            \centering
            \includegraphics[width=\linewidth]{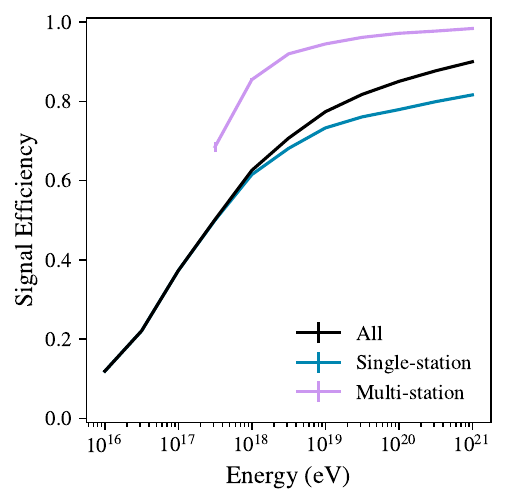}
            \caption{
                \label{fig:single_multistation_efficiency}
                Signal efficiency (i.e.\ fraction of triggered neutrino events that pass the toy event selection described in Section~\ref{sec:calculations}) for events that triggered a single station compared to those that triggered multiple stations across the array. Multi-station results are plotted over the energy range where array-wide simulations were available. The overall signal efficiency to all events is also shown for reference. The signal efficiency has been livetime-averaged over array-wide configurations with all five stations active. In general, analyses are more efficient for events that trigger multiple stations, rather than just one. This increases the overall signal efficiency, particularly at high energies.
            }
        \end{figure}

        The chain of interactions resulting from a single primary neutrino can span the entire detection volume of ARA, creating the potential for multi-station detections of a single event. As shown in Fig.~\ref{fig:multistation}, this class of events becomes increasingly prevalent as the primary neutrino's energy increases (from about $10\%$ of events at $10^{19}$~eV to about $35\%$ at $10^{21}$~eV). This increases the overlap between neighboring stations' detection volumes, resulting in an array-wide acceptance that is sub-additive between stations at high energies. However, this class of events can improve the overall signal efficiency of neutrino searches. Neutrinos which generate triggers in multiple stations are more likely to survive the event selection on at least one station, since they will be observed with different signal strengths and from different directions in each station. This is illustrated in Fig.~\ref{fig:single_multistation_efficiency}, which demonstrates that multi-station detections are responsible for a significant improvement in the overall signal efficiency at high energies. Multi-station triggers also create the potential for a multi-station reconstruction of an event's arrival direction, as discussed in Refs.~\cite{nuradiomc-glaserstudies, nuradiomc-secondaries}. 
        Traditionally, the spacing of large radio arrays has been designed to minimize overlap between stations, but the multi-messenger potential of multi-station triggers may motivate reducing the separation between stations. Future large radio arrays should keep this balance in mind.

        \par
        Multi-station events also increase the complexity of simulations required for large radio arrays, like ARA, RNO-G~\cite{rnog-design}, and IceCube-Gen2 Radio~\cite{gen2-whitepaper}. This is because simulations must capture the state of the full array to accurately estimate its sensitivity, and this state is subject to changes at the station-level. In this analysis, we considered a data-taking period during which individual stations experienced from $2$~up to $9$~stable station-level configurations. Even with only $5$~stations, this results in $50$~unique array-wide configurations. Larger arrays, depending on the stability of their individual stations, will generally need to deal with exponential growth of their array-wide configurations --- each of which will require dedicated simulations. 
        We note that changes to hardware and run-level station settings (such as to waveform readout, trigger windows, and trigger masks) will also motivate the creation of additional station-level configurations.
        As discussed in Section~\ref{sec:simulation}, we have mitigated the compounding increase in required simulations by performing single-station simulations of a shared event library and accounting for multi-station triggers in post-processing. This reduced the overall size of the simulation production required for this study, but also increased its complexity. Future arrays will need to carefully consider the trade-offs between large simulation productions and more computationally intensive post-processing pipelines.

        \par
        Figure~\ref{fig:veffperstation} highlights several important implications for the design of future large radio arrays, reinforcing the results of previous simulation studies~\cite{rno-design-sims, nuradiomc-glaserstudies, gen2-radio-sensitivity}. 
        One can see that deeper stations (e.g.\ the ${\sim}200$~m deep A3) provide more sensitivity at high energies compared to shallower stations (e.g.\ the ${\sim}100$~m deep A1). 
        Generally, this means that future arrays can tune the energies they are most sensitive to, in part, by adjusting the depth at which stations are deployed. 
        Importantly, one can also see that deep stations employing a PA trigger can exceed the low-energy sensitivity of stations using a traditional multiplicity trigger. 
        This increased trigger sensitivity extends to all energies, but is most powerful for low-energy, near-threshold neutrinos.

        \par
        PA triggers also have the advantage of providing good sensitivity while requiring fewer strings to be deployed and, therefore, fewer boreholes to be drilled. This is a significant advantage for the construction of large radio arrays, like IceCube-Gen2 Radio, since it enables a substantial reduction in the total number of boreholes that must be drilled: just one for triggering and any additional strings for improved directional reconstruction~\cite{phased-array-concept}. However, this increased economy of construction also adds additional risk for long-term detector sensitivity. In particular, the loss of a string in the field can diminish reconstruction accuracy (and, therefore, the efficiency of background mitigation) or even the complete sensitivity of a station, if that string is the station's PA trigger. Both of these effects have impacted the performance of A4, which employs a traditional multiplicity trigger but only has $3$~working strings. In particular, Fig.~\ref{fig:veffperstation} shows that this station has a higher energy threshold than those with $4$~working strings to participate in the trigger. For these reasons, it is important that future large radio arrays consider the effect of single-string losses over a portion of the array on their ability to perform directional reconstruction and overall sensitivity. Explorations should also be made into whether redundancies in the trigger chain are possible. Further, this underscores the need for robust pre-deployment testing of hardware and on-station software to minimize the rate of this failure mode. 
        
\section{Conclusion}\label{sec:conclusion}

    \par
    In this study, we assessed the realized sensitivity of ARA's full five-station array to the UHE neutrino flux using an updated simulation framework that incorporates secondary particle interactions for signal generation, as well as new data-driven simulations of the detector over time. This simulation production employed a three-step pipeline using \nuleptonsim{}'s forward-propagation of primaries and secondaries to generate array-wide event libraries, \arasim{}'s single-station detector \& trigger simulations, and a final processing framework to aggregate station-level triggers into an array-wide set of unique primary neutrino triggers. This pipeline was chosen to allow for realistic simulations of the time-dependent detector sensitivity while minimizing the required computational cost.

    \par
    We find that ARA's realized trigger-level acceptance exceeds that of IceCube beyond $2$~EeV. Secondary interactions are responsible for a significant portion of this acceptance, including nearly $40\%$ at $10^{21}$~eV. ARA's large acceptance means that it has the potential to be the most sensitive UHE neutrino detector. Given its exposure from 2013 to 2023, ARA has the leading trigger-level sensitivity to constrain UHE neutrino flux models beyond $4$~EeV. In practice, however, this potential sensitivity is currently not achievable in UHE neutrino searches. To roughly capture the impact of a real UHE neutrino event selection, we applied a simple SNR threshold (to mimic thermal background suppression) and a simple zenith threshold (to mimic suppression of anthropogenic \& cosmic ray backgrounds). The resulting toy analysis-level sensitivity shows that ARA will set the most stringent constraints on UHE neutrino flux models beyond roughly $10$~EeV in a realistic UHE neutrino search. However, the exact energy at which ARA's limit is the most stringent will depend on the details of the specific UHE neutrino search.

    \par
    This study also emphasizes several key considerations for next-generation large radio arrays, like RNO-G and IceCube-Gen2 Radio. While secondary particle interactions increase the sensitivity of detectors, they also result in more complex event waveforms, which should be considered when building analysis pipelines. Otherwise, these potentially powerful events can be lost in event selection. 

    \par
    Multi-station triggers represent a unique event class for multi-messenger astrophysics, but are only present if stations are not too widely spaced. Additionally, these events generally make event selections in UHE neutrino searches more efficient. Future arrays should consider the trade-off between maximizing detector sensitivity and the multi-messenger potential of multi-station events. The existence of multi-station events combined with changing station-level detector configurations also increases the simulation requirements to assess an array's sensitivity. These requirements grow exponentially with the number of stations. Future arrays will need to balance these large simulation productions with more complex processing pipelines. 

    \par
    Additional considerations for detector design include the depth and number of deployed strings, as well as the trigger algorithm. In particular, consideration should be given to detector performance when a single string fails. Single-string PA triggers are economical and effective, but also create additional risk since their failure would result in the loss of an entire station. Robust hardware testing before deployment, along with redundancies in design, should be used to mitigate this risk.

\begin{acknowledgements}
    A.~Bishop, R.~Krebs, M.~Muzio, and A.~Salcedo-Gomez conducted this study and prepared the manuscript. B.~Clark contributed to Appendix~\ref{app:AraSim}.
    \noindent
The ARA Collaboration is grateful for support from the National Science Foundation through Award No.~2013134 and~2310095.
The ARA Collaboration designed, constructed, and now operates the ARA detectors. 
We would like to thank IceCube, and specifically the winterovers, for the support in operating the detector. 
Data processing and calibration, Monte Carlo simulations of the detector and of theoretical models, and data analyses were performed by a large number
of collaboration members, who also discussed and approved the scientific results presented here. 
We are thankful to Antarctic Support Contractor staff, a Leidos unit for field support and enabling our work on the harshest continent. 
We thank the National Science Foundation (NSF) Office of Polar Programs and Physics Division for funding support. 
We further thank the Taiwan National Science Council's Vanguard Program NSC 92-2628-M-002-09 and the Belgian F.R.S.-FNRS and FWO.
K.~Hughes thanks the NSF for support through the Graduate Research Fellowship Program Award No.~1746045. 
A.~Connolly thanks the NSF for Awards No.~1806923 and No.~2209588 and also acknowledges the Ohio Supercomputer Center. 
S.~A.~Wissel thanks the NSF for support through CAREER Award No.~2033500.
A.~Vieregg, C.~Deaconu, N.~Alden, and P.~Windischhofer thank the NSF for Award No.~2411662 and the Research Computing Center at the University of Chicago
for computing resources.
R.~Nichol thanks the Leverhulme Trust for their support. 
K.D.~de~Vries is supported by European Research Council under the European Union's Horizon research and innovation program (Grant Agreement 763 No.~805486). 
D.~Besson, I.~Kravchenko, and D.~Seckel thank the NSF for support through the IceCube EPSCoR Initiative (Award ID No.~2019597). 
M.S.~Muzio thanks the NSF for support through the MPS-Ascend Postdoctoral Fellowship under Award No.~2138121. 
A.~Bishop thanks the Belgian American Education Foundation for their Graduate Fellowship support.

\end{acknowledgements}

\appendix

\section{Comparison to Previous Results\label{app:comp}} 

\begin{figure}[h!]
        \centering
        \includegraphics[width=\linewidth]{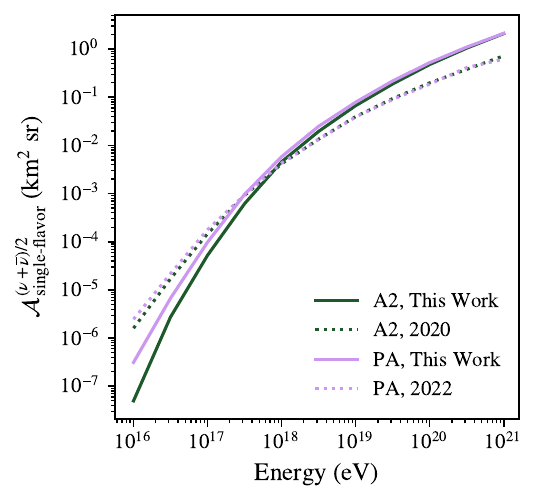}
        \caption{Trigger-level acceptance of A2 and PA calculated in this work compared to results presented for A2 in Ref.~\cite{ara-a23} and PA in Ref.~\cite{ara-pa}, converted into an acceptance.} 
        \label{fig:comp}
    \end{figure}

    In Fig.~\ref{fig:comp}, we show the acceptance (averaged over all configurations' livetimes) for station A2 and the PA calculated in this work compared to previously published results. The enhancement of the acceptance at high energies is due to contributions from secondary interactions, which were not included in previous studies. The suppression of the acceptance at low energies relative to previous studies is due to differences in Earth modeling and the tabulated inelasticity distribution used to determine secondary particle and cascade energies.
    
\section{Angular Sensitivity\label{app:zenith_acceptance}}

    \par
    We calculate the average effective area across bins in elevation angle to characterize ARA's sensitivity to neutrinos arriving from different directions. Since the acceptance is the integral of the effective area $A_\mathrm{eff}(\theta)$ over solid angle, we calculate the average effective area over an elevation angle range $[\theta_\mathrm{lo}, \theta_\mathrm{hi}]$ as

    \begin{align}
        A_\mathrm{eff}(\theta_\mathrm{lo}, \theta_\mathrm{hi}) = \frac{\Delta \mathcal{A}}{\Delta\Omega}~,
    \end{align}
    where $\Delta\Omega = 2\pi ( \sin\theta_\mathrm{hi} - \sin\theta_\mathrm{lo})$ is the solid angle subtended by the bin\footnote{Sine appears in the calculation of solid angle since $\theta$ is an elevation angle, rather than a zenith angle.} and

    \begin{align}
        \Delta\mathcal{A} = \frac{1}{N} \sum_i^{N_\mathrm{trig}} w_i \delta_{\theta_i \in [\theta_\mathrm{lo}, \theta_\mathrm{hi}]}~,
    \end{align}
    where $\theta_i$ is the arrival angle in elevation of event $i$ and $w_i$ is given by \eqref{eq:weights}.

    \par
\begin{figure}
        \centering
        \includegraphics[width=\linewidth]{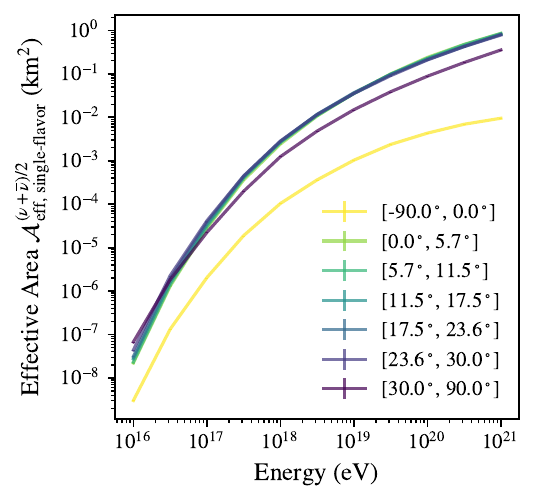}
        \caption{
            Trigger-level effective area of ARA as a function of neutrino arrival direction in terms of elevation angle. The effective area is averaged across the indicated elevation range of the bin. Between $0^\circ$ and $30^\circ$, bins are equal width in solid angle.
            \label{fig:aeff_vs_zenith}
        }
    \end{figure}

    Figure~\ref{fig:aeff_vs_zenith} shows the effective area in equal solid angle bins for elevations between $0^\circ$ and $30^\circ$. Outside of this range, the effective area is calculated in a single bin. As can be seen from Fig.~\ref{fig:aeff_vs_zenith}, most of ARA's sensitivity is to neutrinos arriving within $30^\circ$ above the horizon. These arrival directions are favorable for observation since they provide a long chord through the detection volume that also has a small enough overburden to make survival to that volume probable. Neutrinos arriving from higher elevation angles also have a high survival probability, but ARA's sensitivity to these directions is diminished by the shorter chord lengths along which neutrino interactions could occur in the detection volume. ARA is relatively insensitive to events arriving from below the horizon, and this direction's relative contribution to ARA's overall sensitivity decreases further with energy --- as expected due to the increasing opacity of Earth with energy as the neutrino cross section grows. 
    
\section{\arasim{} \label{app:AraSim}}

    Simulation of the ARA instrument, and Monte Carlo assessment of sensitivity, is achieved with 
    a dedicated C++-based code called \arasim{}.

    The code is publicly available and hosted on GitHub.\footnote{\url{https://github.com/ara-software/AraSim}}
    Core dependencies of \arasim{} include the ROOT package~\cite{ROOT-NIMA-1997}, 
    for class structure and data I/O,
    the Boost libraries, for pointer management, and FFTTools,\footnote{\url{https://github.com/nichol77/libRootFftwWrapper}}
    a FFTW-based package for managing fast Fourier transforms with ROOT wrappers.
    \arasim{} can also optionally link against \araroot{},\footnote{\url{https://github.com/ara-software/AraRoot}} the C++ software package for accessing ARA data and detector information.

    \arasim{} is designed to support end-to-end simulation of neutrino interactions in ARA.
    This includes the physics of the neutrino interaction (``event generation''),
    simulation of the Askaryan electric field, propagation of the field through the ice,
    receipt of the field by the antennas,
    application of the detector response and readout, including noise and trigger,
    and packaging of the simulated event in a data-like format.
        
    \subsection{Event Generation}
        
        \arasim{} supports three ways of generating events.
        First, \arasim{} contains a native Monte Carlo generator 
        of neutrino interactions.
        Second, users may provide a list of particle interactions and locations to be simulated.
        Third, users may directly provide the electric field at each antenna used in the simulation. In the first two cases, the Askaryan electric field from these interactions is generated as discussed in Appendix~\ref{app:askaryan_model}.
        
        \subsubsection{Native Generation of Events}

            \arasim{} was designed to be a backwards simulation of neutrinos.
            The user defines the radius and height of a fiducial cylinder, within 
            which neutrino interaction points are uniformly sampled. The neutrinos are then forced to interact, 
            warranting a weighting scheme to account for sampling bias.
            This unitless weight accounts for the probability that the neutrino 
            survives to its interaction point without an earlier interaction, and is defined as 
            \begin{equation}
              w_{\text{surv}} = \prod_i e^{-(\Delta D_i) / L_{\text{int,i}}}
              \label{eq:arasim_survival_weight}
            \end{equation}
            where the product is over each layer of the Earth model, $\Delta D_i$ is the distance a neutrino travels through that layer, and $L_{\text{int},i}$ is the interaction length of neutrinos in that layer.

            By default, the primary neutrino's flavor ($e$, $\mu$, or $\tau$), and whether it is a particle or anti-particle, is sampled uniformly over all six types of neutrinos.
            The interaction is modeled as a deep inelastic scattering
            via either neutral or charged currents, according to the energy-dependent branching ratios
            from~\cite{connolly-birefringence}.
            Users can change which neutrino flavors and interaction types are sampled for more specific studies.
            
            Several approximations are made to simplify the simulation of interaction processes.
            For example, \arasim{} simulates the decay of tau leptons only if the decay's energy deposition is greater than the cascade accompanying its production. In this case, only the signal generated by the decay is simulated.
            Additionally, Glashow resonance interactions are neglected by \arasim{}'s native event generator.
    
        \subsubsection{Event List Read-In\label{sec:eventlist}}

            A user may elect to simulate neutrino interactions with another framework but use \arasim{} to simulate the electric field from those interactions and the detector response.
            To do this, an event generator (\nuleptonsim{}, in this study~\cite{Cummings:2023iuw}) is used to create a text file containing each interacting particle's flavor and whether it is an anti-particle; the interaction's type, inelasticity, location, and direction; and the primary neutrino's tracking ID and energy.
            \arasim{} uses the interaction's parameters to calculate the electric field from the interaction and processes the event through the
            remainder of the simulation steps (field propagation, detector simulation, etc.).
            If multiple events in the file list have the same primary neutrino tracking ID, these interactions are simulated as part of the same event. 
            \arasim{} combines the signals from each of these interactions in a single voltage trace, delayed according to their relative interaction and propagation times.
            The resulting voltage trace is automatically lengthened to accommodate all of the contributing interactions.
        
        \subsubsection{Electric Field Read-In}

            A user may also provide a series of files to \arasim{} describing the Cartesian components of the electric field at each antenna as a function of time. 
            This allows users to import results from other simulation frameworks, such as \texttt{FAERIE}~\cite{FAERIE}, \texttt{CoREAS}~\cite{COREAS}, and \texttt{Corsika~8}~\cite{corsika8} into \arasim{}.
            This allows a user to simulate the response of the
            ARA detector to non-neutrino physics signals,
            such as cosmic rays and beyond the Standard Model (BSM) processes, among others.

    \subsection{Askaryan Emission Model} \label{app:askaryan_model}

            \arasim{} supports two Askaryan emission models.
            They are described in detail elsewhere~\cite{hong2014},
            so we only briefly recount them here.
            The first is a ``time domain'' model, where the electric field is found by time-differentiating the vector potential for the particle shower.
            The vector potential $\mathbf{A}(\theta, t)$
            under a far-field assumption has the following form:
            \begin{equation}
            \mathbf{A}(\theta, t) \propto\int_{-\infty}^{\infty} dz' Q(z') \cdot F_p\!\left(t, z'\right)
            \end{equation}
            where $z'$ is the shower depth in meters, $Q(z')$
            is the shower charge excess as a function of depth,
            $t$ is time, and $\theta$ is the angle between the shower axis
            and the RF signal propagation direction (viewing angle).            
            $F_p$ is the ``shower form factor''~\cite{AlvarezMuniz2012}, that 
            governs the shape of the radiated signal.
            The parameters of $F(p)$ are extracted from fits
            to the vector potentials from full simulations of 
            showers in media with ZHAireS~\cite{2011PhRvD..84j3003A}.
            Note that \arasim{} performs a full integral over the longitudinal
            profile of the shower, so that this model generates
            radiated signals with proper phase information.
            This ``time-domain'' model, with correct phases,
            is the one used in this paper.

            The second Askaryan model is a ``frequency-domain'' model,
            which is an implementation of the model in~\cite{showerenergy},
            and returns only the spectral amplitude as a function of frequency $f$:
            \begin{align}
              E(f) \propto \frac{f}{1150~\text{MHz}}\times \text{exp}\left[ - \frac{1}{2} \left( \frac{f}{1~\text{GHz}} \times \frac{\Omega}{2.2^{\circ}} \right)\right]
              \label{eq:alvarez-muniz}
            \end{align}
            The angular width of the cone of Askaryan radiation, $\Omega$, then follows the equations in Ref.~\cite{had-cone} for hadronic cascades and Ref~\cite{em-cone-lpm} for electromagnetic cascades. 
            Since this model only generates emission in the frequency
            domain, it does not account for the
            phases of the signal, and it therefore exaggerates the
            impulsiveness of the emission.
            This comparatively simple frequency-domain model is useful for comparisons to other simulation frameworks.

            In both emission models, the length of the electromagnetic particle shower is extended
            if the shower energy is greater than $10^{15}$~eV due to the Landau–Pomeranchuk–Migdal (LPM) effect following~\cite{em-cone-lpm}.
            The LPM effect changes $F(p)$ in the time-domain model, while in the frequency-domain model it alters $\Omega$.

    \subsection{Modeling the Environment}

        \par
        After signals are generated, they must be propagated from the interaction vertex position to each of the station's channels. The properties of the Antarctic glacial ice have important effects on the propagation of radio waves, namely refractive, attenuation, and birefringence effects, all of which \arasim{} can simulate. Below, we describe how these effects are modeled, though birefringence effects~\cite{connolly-birefringence,SalcedoGomez:2024rX} are not considered in this study.
    
        \subsubsection{Index of Refraction}
            \par
            A refractive index model for the ice near ARA is required to determine the path of signal rays and their relative arrival times at each antenna.
            At leading order, the index of refraction $n$ depends only on depth from the surface $z$, which arises due to compactification
            of the glacial ice as new snow falls annually.
            The medium transitions from cold ($\sim -55~^\circ$C) loose snow at the surface, to relatively warm ($\sim -10~^\circ$C) ice at the bottom of the glacier~\cite{icetemp-data}.
            The implemented model for $n(z)$, which arises from ice-density arguments~\cite{robin1969-radio-echo}, takes the functional form
            of an exponential:

            \begin{align}
                n(z) = n_\mathrm{ice} - (n_\mathrm{ice}-n_\mathrm{firn}) e^{-z/\Lambda}~,
            \end{align}
            where $n_\mathrm{ice}$ and $n_\mathrm{firn}$ are the ice's refractive indices in bulk and firn regions, respectively, and $\Lambda$ is the length scale of the transition between these two regions.
            Other functional forms for $n(z)$ are available in the literature (e.g.~\cite{Ali:2024zgx}), but are not currently supported. 

            \par
            The values for $n_\mathrm{ice}$, $n_\mathrm{firn}$, and $\Lambda$
            must be derived from data.
            Several are available in \arasim{}, including those from the RICE experiment~\cite{Kravchenko-Besson-Meyers-2004} and
            previous ARA studies~\cite{ara-a23, ara-pa}.
            For this paper, we adopted the model used in~\cite{ara-pa} with $n_\mathrm{ice} = 1.780$, $n_\mathrm{firn} = 1.326$, and $\Lambda = 0.0202^{-1}$~m~$\simeq 49.505$~m.

            \par
            Because the index of refraction depends on depth,
            Snell's law dictates that light paths follow curved, rather than rectilinear trajectories.
            This makes finding the paths non-trivial.
            To determine the ray paths, \arasim{} uses a numerical ray tracer that solves for the characteristics of the eikonal equation,
            \begin{align}
                \lvert \nabla T(\mathbf{r}) \rvert = n(\mathbf{r})~,
            \end{align}
            via a 4th-order Runge-Kutta integration, where $T$ is the propagation time, $n$ is the index of refraction, and $\mathbf{r}$ is a position in the ice. This numerical integration allows reflections at both the air-ice and ice-bedrock boundaries to be incorporated. 

            \par
            Ray tracing is performed in two modes, depending on whether an initial-value or boundary-value problem is being solved. To solve an initial-value problem, an initial position and launch angle are provided, and the ray is determined via pure forward integration. More commonly, though, only an initial position (the interaction vertex location) and a final position (the antenna location) are provided. In this case, \arasim{} follows a two-step approach. The first ray-trace solution (the shortest, ``direct'', ray) is determined by calculating the launch angle via the semi-analytic method described in~\cite{ara-reco-prototypestation}. Additional ray-trace solutions (in particular, the second-shortest, ``refracted'', ray) are found by determining other launch angles that minimize the distance of the forward-integrated ray to the target location via a root-finding algorithm. This hybrid approach is more computationally efficient compared to traditional trial-and-error methods.
            The ray tracer tracks the emission and receipt angles, 
            the path through the ice, and the travel time.
    
        \subsubsection{Ice Attenuation}
            Radio signals propagating in ice are attenuated over $\mathcal{O}(1)$~\text{km}~length scales~\cite{SPice-attenuation}. \arasim{} has several, increasingly detailed, implementations of the attenuation length. 

            \par
            For this study, we used the most detailed implementation, which calculates the attenuation length as a function of both depth and frequency. The attenuation length's dependence on depth is due to the depth-dependence of the ice temperature, which in \arasim{} is parametrized via a cubic polynomial

            \begin{align}
                t(z) = az^3 + bz^2 + cz + d~,
            \end{align}
            where the $t$ is the temperature in degrees Celsius, $z$ is the distance from the ice surface, and the coefficients were determined from a fit to data from AMANDA and IceCube~\cite{icetemp-data}: $a=1.83415\times 10^{-9}$~$^\circ$C/m$^3$, $b=-1.59061\times 10^{-8}$~$^\circ$C/m$^2$, $c=2.67687\times 10^{-3}$~$^\circ$C/m, and $d=-51.0696$~$^\circ$C. 

            \par
            Given an ice temperature, the attenuation length at three reference frequencies is modeled via a quadratic polynomial in temperature:

        \begin{align}\label{eq:reference_attenuation_lengths}
                \ln\left(\frac{1~\text{m}}{\lambda_\mathrm{att}(f_i,t)}\right) = \alpha_i t^2 + \beta_i t + \gamma_i~,
            \end{align}
            where the coefficients, given in Table~\ref{tab:attenuation_length_coefficients}, were obtained by fitting data from~\cite{bogorodsky2012radioglaciology}. The final attenuation length is then obtained by linearly interpolating in log-frequency space between these reference frequencies:

            \begin{align}
                \ln\left(\frac{1~\text{m}}{\lambda_\mathrm{att}(f)}\right) =~ &\ln\left(\frac{1~\text{m}}{\lambda_\mathrm{att}(f_i)}\right) \nonumber \\ &+ \frac{\ln(f/f_i)}{\ln(f_{i+1}/f_i)} \ln\left(\frac{\lambda_\mathrm{att}(f_i)}{\lambda_\mathrm{att}(f_{i+1})}\right) ~,
            \end{align}
            where $i$ is such that $f_i \leq f < f_{i+1}$ and the dependence on temperature and depth has been suppressed for simplicity.

            \begin{table}[]
                \centering
                \begin{tabularx}{\linewidth}{c S S S}
                    \toprule
                    {$f_i$} & {$\alpha_i$ ($^\circ$C$^{-2}$)} & {$\beta_i$ ($^\circ$C$^{-1}$)} & {$\gamma_i$} \\
                    \midrule
                    $0.1$~MHz & -0.000884 & 0.026709 & -6.74890\\
                    $1$~GHz & -0.001773 & -0.070927 & -6.22121 \\
                    $3.16$~GHz & -0.000332 & -0.002213 & -4.09468 \\
                    \bottomrule
                \end{tabularx}
                \caption{Coefficients for the quadratic function~\eqref{eq:reference_attenuation_lengths} used to model the attenuation length at three reference energies: $f_0=0.1$~MHz, $f_1=1$~GHz, and $f_2=3.16$~GHz.
                \label{tab:attenuation_length_coefficients}}
            \end{table}

            To calculate the total attenuation factor experienced by a ray, \arasim{} calculates the product of attenuation factors in small steps along the ray path

            \begin{align}
                f_\mathrm{att}(f) = \prod_i \exp\left(-\frac{L_i}{\lambda_\mathrm{att}(f,z_i)}\right)~,
            \end{align}
            where $L_i$ is the length of the $i$th ray step and $z_i$ is the average depth in that step. This factor is then applied as $E_\mathrm{rec}(f) = f_\mathrm{att}(f) E_\mathrm{emit}(f)$, where $E_\mathrm{rec}$ is the frequency-space electric field amplitude arriving at the antenna and $E_\mathrm{emit}$ is its amplitude at the emission point.

            \subsubsection{Birefringence}
            
            \par
            In addition, radio propagation in South Pole ice may be affected by anisotropies in the dielectric properties of the medium. In an anisotropic medium, the propagation speed depends on the local direction of propagation and the polarization of the signal. As a result, an incident electric field decomposes into two orthogonal propagation eigenstates, each of which experiences a different refractive index.

            \par
            The model implemented in \arasim{} follows the approach of~\cite{connolly-birefringence}, in which the anisotropy is described by three depth-dependent principal (refractive) indices, $(n_\alpha, n_\beta, n_\gamma)$, obtained from Ref.~\cite{voigt2017-spice} and defined along orthogonal axes of the medium. In this framework, the local dielectric structure is represented by an indicatrix whose semi-axes are set by these three principal indices. For a signal propagating with local wave vector $\mathbf{k}$, the plane normal to $\mathbf{k}$ intersects the indicatrix in an ellipse. The principal axes of this ellipse define two local displacement-field eigenvectors, $\mathbf{D}_1$ and $\mathbf{D}_2$, and the corresponding semi-axis lengths determine the refractive indices $n_1$ and $n_2$ experienced by the two eigenstates. 

            \par
            The ray path is determined by the standard ray tracing described above (i.e.\ the trajectory is calculated without birefringence corrections) and then subdivided into segments over which the propagation direction and local index of refraction are assumed to be constant. At each step, the local eigenstates and their associated refractive indices are calculated from the current propagation direction and the local principal indices. The accumulated time delay between the two eigenstates at the receiver is then

            \begin{align}
                \Delta T = \sum_i \frac{\Delta \ell_i}{c}\left(n_{2,i}-n_{1,i}\right),
            \end{align}
            where $\Delta \ell_i$ is the path length of segment $i$, and $n_{1,i}$ and $n_{2,i}$ are the effective refractive indices of the two eigenstates on that segment.

            \par
            Additionally, this model assumes that the local eigenstates evolve adiabatically along the ray path, so that the signal polarization continuously follows the rotation of the eigenstates. Therefore, birefringence modifies both the relative arrival times and the polarization of the received signals. 
            
    \subsection{Modeling the Detector}\label{app:detector_modeling}
    
        After an electric field arrives at the station,
        \arasim{} must model the response of the detector.
    
        \subsubsection{Array and Station Layout}
        
            \arasim{} can be used to simulate up to $37$~identical ARA stations in a hexagonal layout with user-defined inter-station spacing.
            This capability supported the original design studies for the array.
            Simulation of ARA's Testbed, the five ``traditional'' quad-string stations, is also supported.
            The Testbed is composed of 10 antennas deployed on 5 strings to a depth of at most 30~m~\cite{ara-design}.
            Typically, though, \arasim{} is used in a single-station simulation mode, where each station (A1--A5 and PA) is simulated using parameters imported from \araroot{}. These parameters include the station's calibrated antenna positions and other time-dependent properties, such as the exclusion of channels from the triggering logic.
            
            The PA detector, in particular, has three distinct simulation modes,
            representing the time evolution of the sub-detector.
            The first mode simulates the PA string as it was deployed in 2018,
            where it operated with seven VPol antennas and two HPol antennas.
            The second mode simulates the PA in 2019 when one VPol antenna from A5 was split and read out by both the A5 and PA DAQ boards. 
            The third mode simulates the PA after 6 more VPol antennas from A5 were connected to the PA DAQ during the 2019--2020 season to aid in event reconstruction.
            
        \subsubsection{Antenna Gain Modeling \label{app:antenna}}

        After the electric field is propagated to the antenna location, \arasim{} converts the electric field into the voltage response of the antenna by convolving the signal with the antenna model in the frequency domain.

        For each ray solution, the electric field at the antenna is expressed in a local spherical basis $(\hat{r},\hat{\theta},\hat{\phi})$, where $\hat{r}$ points along the propagation direction of the incoming plane wave. VPol antennas are sensitive to the $\hat{\theta}$ component of the field, while HPol antennas are sensitive to the $\hat{\phi}$ component.
        
        The signal is then convolved with the antenna model frequency-by-frequency.
        For each frequency component $f$, the amplitude is scaled by the antenna vector effective height, while the phase is shifted according to the antenna phase response.
        In \arasim{}, the magnitude of the vector effective height is derived from the realized antenna gain according to
        \begin{equation}
        |\mathbf{h}_{\rm eff}(f,\theta,\phi)|
        =
        \sqrt{
        \frac{G_r(f_{s},\theta,\phi)\, c^2\, Z_r}
        {4\pi\, n_{\rm eff}\, f^2\, Z_0}
        }, \label{eq:heff}
        \end{equation}
        where $G_r(f_s,\theta,\phi)$ is the realized gain, $f_s = n_{\rm eff} f / n$ is the medium-scaled frequency (see discussion below), $n_{\rm eff}$ is the effective refractive index of the antenna in its borehole, $n$ is the refractive index in which the frequency-domain antenna gain was measured, $c$ is the speed of light in vacuum, $Z_0$ is the impedance of free space, and $Z_r$ is the receiver impedance.
        The vector effective height is the product of the magnitude with the
        relevant unit vector.
        For example, for a VPol antenna this would be $\mathbf{h}_{\rm eff} = |\mathbf{h}_{\rm eff}| \hat{\theta}$.
        The corresponding frequency-domain voltage at the antenna can then be written as
        \begin{equation}
        V_\mathrm{ant}(f) = \mathbf{E}(f) \cdot \mathbf{h}_{\rm eff}(f,\theta,\phi)\, e^{i\Phi_\mathrm{ant}(f,\theta,\phi)},
        \end{equation}
        where $\mathbf{E}(f)$ is the incident electric-field and $\Phi_\mathrm{ant}(f,\theta,\phi)$ is the antenna phase.

        For the gain of the antenna, \arasim{} supports several antenna-response model types, including models derived from electromagnetic simulations, from in-situ measurements, and anechoic chamber measurements. Separate responses are used for the bottom VPol, top VPol, and HPol antennas. For each antenna type, \arasim{} uses a realized gain
        $G_r(f,\theta,\phi)$
        and phase
        $\Phi_\mathrm{ant}(f,\theta,\phi)$,
        where $f$ is the signal frequency and $(\theta,\phi)$ specifies the signal arrival direction in the antenna coordinate system.
        
        For this work, we use anechoic chamber measurements of ARA antenna responses taken at the University of Kansas~\cite{KU-antennas}. Since these responses were measured in air, while the detector operates in ice, an additional translation step is required before they can be applied in simulation. To account for this difference, \arasim{} translates the antenna response from the medium of the measurement to the local detector medium through a refractive-index scaling,
        \begin{equation}
        f_{\rm s} = f\cdot \frac{n_{\rm target}}{n_{\rm source}},
        \end{equation}
        where $n_{\rm source}$ is the refractive index of the medium in which the antenna response was measured, and $n_{\rm target}$ is the refractive index of the medium in which the antenna is evaluated in simulation. This scaling accounts, at leading order, for the shift in the antenna's resonant frequency between media.
        
        The antenna model also includes information of impedance-mismatch of the antenna and its load through the standing-wave ratio (SWR), which is converted into a transmission coefficient and separately incorporated into the detector response:
        \begin{equation}
        \Gamma = \frac{\mathrm{SWR}-1}{\mathrm{SWR}+1},
        \qquad
        T = \sqrt{1-\Gamma^2},
        \end{equation}
        where $\Gamma$ is the voltage reflection coefficient and $T$ is the corresponding transmission factor.
        
        \subsubsection{Electronics Gain Modeling}

\begin{figure}
                \centering
                \includegraphics[width=\linewidth]{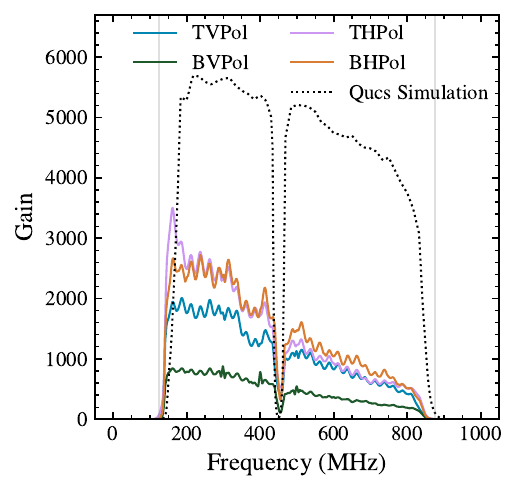}
                \caption{
                    \label{fig:gain_model}
                    Comparison between (linear) gain models of the system response in \arasim{}, as a function of frequency. The previously used model, based on a \texttt{Qucs} simulation of the system (dotted), is compared to the data-driven models (solid lines, this work) used in this work, for the example of String 1 of A3 in configuration 1. Data-driven models are presented for each channel: top VPol (TVPol), bottom VPol (BVPol), top HPol (THPol), and bottom HPol (BHPol). 
                    }
            \end{figure}

        After the antenna, \arasim{} models the response
        of the detector's amplifiers and filters.
        Similar to the antenna response, the detector electronics response is described through a frequency-dependent electronics gain $G_\mathrm{elec}(f)$
        and phase
        $\phi_\mathrm{elec}(f)$. \arasim{} supports either simulated electronics response models (e.g.\ \texttt{Qucs} simulations of the system) or in-situ, station- and configuration-dependent responses extracted from forced trigger data,\footnote{ARA stations use a supplementary $1$~Hz ``forced trigger'' to monitor the ambient background environment.} as shown in Figure~\ref{fig:gain_model}.

        For each channel, the electronics response is applied in the frequency domain. The voltage spectrum before the electronics chain, $V_{\rm ant}(f)$, is converted to the voltage spectrum after the electronics response as
        \begin{equation}
        V_{\rm elec}(f) = V_{\rm ant}(f)\, G_{\rm elec}(f)\, e^{-i\phi_{\rm elec}(f)},
        \end{equation}
        where $G_{\rm elec}(f)$ and $\phi_{\rm elec}(f)$ are the frequency-dependent electronics gain and phase response for the corresponding channel.

        To determine the in-situ gain model, forced trigger waveforms are Fourier transformed to obtain the measured voltage spectrum for each station, configuration, and channel. This measured spectrum, $H_{\mathrm{meas}}(f)$, contains the thermal noise entering through the antenna convolved with the electronics gain. The expected thermal contribution at the antenna, $H_{\mathrm{theory}}(f)$, is calculated from the antenna transmission coefficient and total noise temperature. The electronics gain model is then obtained by dividing the measured spectrum by the expected thermal spectrum,
        \begin{equation}
            G_{\mathrm{elec}}(f) = \frac{H_{\mathrm{meas}}(f)}{H_{\mathrm{theory}}(f)}.
        \end{equation}
        \arasim{} recalculates the gain model to ensure it is consistent with the user-selected noise model (discussed in the following section) and antenna gain model.

        \subsubsection{Noise Modeling}

            \par
            After the antenna and electronics response have been applied, noise is added to the signal.
            In frequency space, thermal noise has spectral amplitudes drawn from a Rayleigh distribution. A noise model is therefore fully characterized by the scale parameter $\sigma(f)$ of the Rayleigh distribution at each frequency $f$.
            
            \arasim{} has two noise model implementations.
            The first is an idealized noise model,  
            which assumes white thermal noise across all frequencies.
            This is unrealistic, but it is useful for basic tests 
            of the simulation package or 
            to simplify assumptions for comparisons to other frameworks.
            In this idealized model, the Rayleigh distribution scale factor is determined assuming flat Johnson-Nyquist noise.
            For this purpose, we use a load resistance of $Z=50$~$\Omega$ and a total noise temperature of $T=325$~K to reflect both the $230$~K average ice temperature and the $95$~K noise temperature of the low noise amplifier (LNA). 

            The second model, which is data-driven, is used in this study.
            To determine $\sigma(f)$ from data, the distribution of spectral amplitudes in frequency bins is generated from forced-trigger data.
            The resulting distribution is fit to a Rayleigh distribution to determine the scale factor, $\sigma_{r,a}(f)$, for each frequency bin and channel, $a$, on a station for that run, $r$. These values are averaged over all the runs in a livetime configuration to determine the final scale factor model $\sigma_{s,c,a}(f)$ for that channel $a$, livetime configuration $c$, and station $s$.
            
            \par
            Figure~\ref{fig:noise_model} shows a comparison of the noise models derived from in-situ data. These are shown after the system response has been applied.

            \arasim{} generates noise waveforms by sampling the frequency-space voltage amplitude from a Rayleigh distribution with scale $\sigma(f)$ given by the chosen model and phase sampled uniformly between $0$ and $2\pi$. The simulated noise trace is obtained by transforming the voltage back to the time-domain.
        
        \subsubsection{Traditional Trigger Simulation}

        Traditional ARA stations, A1--A5, use a
        multiplicity trigger.
        The trigger condition requires that impulsive power be observed independently in multiple channels
        within a coincidence window.
        In \arasim{}, the voltage waveform in each channel after the electronics response $V_{elec}(t)$ is first passed through a model of the tunnel diode, which acts as a power integrator. The diode output can be written as
        \begin{equation}
        D_i(t) = |V_{\mathrm{elec}, i}(t)|^2 * h_{\rm diode}(t),
        \end{equation}
        where $V_{\mathrm{elect}, i}(t)$ denotes the voltage waveform in channel $i$, $h_{\rm diode}(t)$ is the tunnel-diode response function, and $*$ denotes a convolution. A channel is considered to have triggered when the diode output exceeds a threshold $
        D_i(t) > D_{\rm th}$.
        
        The detector triggers when a configurable number, typically three, of like-polarization channels satisfy this condition within a coincidence window $\Delta t_{\rm trig}$.
        Physically, this reflects the expectation that a true impulsive plane-wave signal should appear across several antennas within a window, given roughly by the light-like travel time. The coincidence window is taken to be $\Delta t_{\rm trig} \simeq 170~{\rm ns}$, which corresponds to roughly the maximum time required for a wavefront to traverse the detector.
        
        \subsubsection{Phased Array Trigger Simulation}

        For the phased array station, the trigger is designed to improve sensitivity to low-amplitude signals using coherent beamforming. In a beamformed trigger, signals from $N$ closely spaced antennas are delayed and summed so that a plane wave arriving from a given direction adds coherently. For a coherent signal, the summed amplitude scales as $N$, while the RMS of incoherent thermal noise scales as $\sqrt{N}$, so that the $\mathrm{SNR} \propto \sqrt{N}$.

        In \arasim{}, this trigger is represented through a trigger efficiency-based proxy rather than a full hardware-level simulation of beamforming. The simulation calculates the signal-only SNR in a short window in a reference VPol channel, chosen in the current implementation to be the top-most VPol antenna.

        This SNR is converted to the equivalent SNR for a signal arriving from the direction of the local calibration pulser, $\theta_\mathrm{CP}$. This is done by scaling the measured SNR by the ratio of the PA's $50\%$ trigger efficiency SNR to signals from $\theta_\mathrm{CP}$ to that from the signal arrival direction $\theta$,
        \begin{equation}
        {\rm SNR}_{\rm eff} = {\rm SNR}_{\rm meas}\, \frac{{\rm SNR}_{50\%}(\theta_\mathrm{CP})}{{\rm SNR}_{50\%}(\theta)},
        \end{equation}
        where ${\rm SNR}_{50\%}(\theta_\mathrm{CP})=2$ and ${\rm SNR}_{50\%}(\theta)$ is the angular-response described in~\cite{ara-pa-design}. The resulting effective SNR is mapped onto a trigger-efficiency curve $ \epsilon_{\rm PA}({\rm SNR}_{\rm eff})$, determined from measurements. The PA then triggers randomly with a probability equal to $\epsilon_{\rm PA}({\rm SNR}_{\rm eff})$. 

        \subsubsection{Waveform Generation}

        \par
        \arasim{} saves a waveform for every successful trigger of the detector in simulation. Each trigger has an associated trigger time about which a readout window is defined. The size of the readout window, and how it is centered relative to the trigger time, is a configurable parameter that is set to match the settings of the real detector in the desired time period. The final voltage trace (after inclusion of signal, noise, and all detector effects) is then saved in the readout window for each channel. As discussed in Appendix~\ref{app:event_readout}, the saved simulated waveform is formatted to match real calibrated data.
    
    \subsection{Event Readout}\label{app:event_readout}

        \arasim{} saves all information using the ROOT file format.
        ROOT organizes files into ``trees,''
        where each tree contains ``branches'' that store individual variables or arrays.
        Each entry in a tree corresponds to a single event,
        allowing efficient, column-oriented access to large datasets
        without loading the entire file into memory.
        This structure makes it straightforward to selectively read
        only the branches relevant to a given analysis,
        which is particularly useful when processing large files.

        \subsubsection{Simulation Settings (\texttt{AraTree})}
        
        The simulation configuration is stored in the \texttt{AraTree}, which contains a single entry describing the detector configuration and other simulation conditions. This includes the detector geometry, the ice model, trigger configuration, and other simulation settings.
        In \arasim{}, \texttt{AraTree} is constructed with branches containing objects used to initialize the simulation:
        \begin{itemize}
            \item \texttt{Detector}, which defines the antenna layout and hardware responses,
            \item \texttt{IceModel}, which specifies the depth-dependent index of refraction and attenuation properties of the ice, 
            \item \texttt{Trigger}, which contains the trigger type and thresholds,
            \item \texttt{Settings}, which controls all simulation parameters,
            \item \texttt{Spectra}, which defines the neutrino flux model.
        \end{itemize}
        Since these objects are written once per simulation, \texttt{AraTree} serves as a record of the simulation configuration.
        
        \subsubsection{Simulation Results (\texttt{AraTree2})}
        
        The event-level simulation results are stored in \texttt{AraTree2}, which contains one entry per simulated neutrino event, or only triggered events upon request. Each entry includes both the generated event properties and the detector response. Specifically, \texttt{AraTree2} stores:
        \begin{itemize}
            \item The \texttt{Event} object, which carries the neutrino interaction properties such as vertex position, direction, energy, and interaction type,
            \item The \texttt{Report} object, which contains the detector response to that event. 
        \end{itemize}

        The \texttt{Report} class holds information from all antennas and ray solutions. For each interaction and antenna, it stores quantities such as ray-tracing solutions between the vertex and antenna, propagation distances and attenuation, viewing and launch angles, polarization vectors, and the frequency-domain and time-domain voltage responses. If enabled, \arasim{} can store intermediate quantities, such as electric fields and frequency-domain signals prior to convolution with the detector response.
        
        \subsubsection{Results Mimicking Data (\texttt{eventTree})}
        
        In addition to simulation-specific output, \arasim{} produces a data-like representation of each event in \texttt{eventTree}. This tree is designed to mimic the structure of real ARA data and can be directly processed using analysis and reconstruction tools for ARA data.
        In particular, the \texttt{eventTree} contains a \texttt{UsefulAtriStationEvent}, which is the proprietary ARA data format,
        from \araroot{}, containing calibrated waveforms for each channel, 
        metadata, and an event weight corresponding to the simulated neutrino.
        These objects hold the final voltage waveforms after all detector effects have been applied, including signal propagation, antenna response, electronics response, and noise generation.
        
\bibliographystyle{apsrev4-2}
\bibliography{main}

\end{document}